\documentclass[11pt]{article}
\usepackage{jheppub}
\pdfoutput=1
\usepackage{amsmath}
\usepackage{epsfig}
\usepackage{amssymb}
\usepackage{graphicx}
\usepackage{epstopdf}
\usepackage{pdfsync}
\usepackage{shuffle}
\usepackage{subfigure}
\usepackage{stix}  
\usepackage[table]{xcolor}
\usepackage{hyperref}
\hypersetup{colorlinks=true,linkcolor=violet,anchorcolor=green,citecolor=purple,filecolor=black,menucolor=black,urlcolor=purple}
\preprint{ }
\title{BCJ, worldsheet quantum algebra and KZ equations}
 \author[a]{Chih-Hao Fu}
 \author[b]{, Yihong Wang}

  \affiliation{${}^{a}$ School of Physics and Information Technology,
  Shaanxi Normal University,\\
  No.620 West Chang'an Avenue, Xi'an 710119, P.R. China.}
 \affiliation{${}^{b}$Department of Physics, National Taiwan University,\\
 No.1 Sec.4 Roosevelt Road Taipei 10617,Taiwan (R.O.C.).}

 \emailAdd{chihhaofu@snnu.edu.cn}
 \emailAdd{yihongwang@phys.ntu.edu.tw}

\date{\today} \abstract{
We exploit the correspondence between twisted homology and quantum group to construct an algebra explanation of the open string kinematic numerator. In this setting the representation depends on string modes, and therefore the cohomology content of the numerator, as well as the location of the punctures. We show that quantum group root system thus identified helps determine the Casimir appears in the Knizhnik-Zamolodchikov connection, which can be used to relate representations associated with different puncture locations.
}

\begin{document}
\maketitle \flushbottom


\section{Introduction}

In their original formulation, Bern, Carrasco and Johansson (BCJ)
proposed a new symmetry feature between the colour and kinematic
factors of Yang-Mills amplitude at generic $n$ points and helicity
configurations \cite{Bern:2008qj,Bern:2010ue}. Different from the 
standard Feynman rules construct
that defines the amplitude, the new formulation states that the Yang-Mills
amplitude is expressible as a double copy. 
\begin{equation}
\mathcal{M}_{YM}=\sum_{\text{cubic graph }i}\,\frac{c_{i}\,n_{i}}{D_{i}}
\label{eq:double-copy-1}
\end{equation}
The above formula resembles the structure-wise simpler bi-adjoint
$\phi^{3}$ theory in the sense that the amplitude is organised as
a sum of all possible cubic graphs divided by appropriate propagators
$D_{i}$, except that one copy of the gauge group contribution is
replaced by a momentum dependent factor known as the kinematic numerator
$n_{i}$. The $n_{i}$'s are assumed to satisfy all the Jacobi and
anti-symmetry relations as the cubic graphs prior to the replacement,
for example at four points the three numerators associated with each
channel sum up to zero,
\begin{equation}
n_{s}+n_{t}+n_{u}=0\,\longleftrightarrow\,f^{a,b}{}_{\sigma}f^{\sigma,c}{}_{d}+\text{cyclic perm of }a,b,c=0
\end{equation}
and this correspondence between colour and kinematics is assumed to
hold for all loop levels provided we integrate (\ref{eq:double-copy-1})
over loop momenta. In addition it was conjectured when both sets of
numerators of a $\phi^{3}$ are replaced by kinematic numerators,
the result correctly reproduces the amplitude of gravitons coupled
to dilatons and $B$ fields. The BCJ double copy structure has detectable
consequences on the whole scattering amplitude. In physically
realised theories the gauge group is $SU(N)$ and its dependence factorises,
whereas each ordering of the Chan-Paton factor defines a partial (or
\textit{colour-ordered}) amplitude. At tree level it was realised
that the number of independent kinematic numerators, counted each
as a degree of freedom, is fewer than the number of partial amplitudes.
As a consequence the partial amplitude needs to satisfy the following
BCJ amplitude identity.
\begin{equation}
k_{1}\cdot k_{2}A(1,2,3,\dots,n)+(k_{1}+k_{3})\cdot k_{2}A(1,3,2,\dots,n)+\dots+(k_{1}+\dots+k_{n-1})\cdot k_{2}A(1,3,\dots n-1,2,n)=0.\label{eq:bcj-amp-id}
\end{equation}
The aforementioned duality structure originally observed in Yang-Mills
amplitudes was later realised to be a common feature of a wide range
of quantum field theories including a variety of supersymmetric gravity
and gauge theories \cite{Bern:2008qj,Bern:2010ue,Bern:2011rj,Bern:2009kd,Bern:2014sna,Bern:2013uka,Johansson:2014zca,Chiodaroli:2014xia,Chiodaroli:2015wal,Ben-Shahar:2018uie,Anastasiou:2017nsz,Carrasco:2012ca,Anastasiou:2015vba,Damgaard:2012fb,Johansson:2017srf,Johansson:2018ues} as well as effective field theories such as nonlinear
sigma model \cite{Chen:2013fya}, Dirac-Born-Infeld, special-Galilean theory \cite{Cachazo:2014xea}, and perhaps
most importantly the physical theory that incorporates 
fermions \cite{Johansson:2014zca,Johansson:2015oia,Johansson:2019dnu}. In
addition it is known to present in the classical solution of Einstein's
equations \cite{Monteiro:2014cda,Luna:2016hge}. The generic feature is a double copy of colour and/or kinematics
whereas replacing one or both copies reproduces amplitude of one
theory from another. The compatibility of gauge invariance among these
theories with double copy has been shown to imply supersymmetry and
diffeomorphism invariance and is proved to be a powerful tool in the
understanding of analytic behaviour of generic quantum field theory.
For more details we refer the readers to the very readable recent
review \cite{Bern:2019prr} and the references therein. It is perhaps
worth emphasise that the interest in further investigation of BCJ
duality is more than an academic one. Indeed, some of the most cutting-edge
higher loop level corrections to the gluon scattering amplitudes were
obtained through double copy 
\cite{Bern:2010ue,Bern:2012uf,Bern:2013qca,Bern:2014sna,Mafra:2014oia,Mafra:2014gja,Mafra:2015mja,Boels:2013bi,Carrasco:2011mn,Bjerrum-Bohr:2013iza,Carrasco:2012ca,Bern:2013yya,Nohle:2013bfa,Ochirov:2013xba,Chiodaroli:2013upa,Chiodaroli:2014xia,Yuan:2012rg,Mogull:2015adi,He:2015wgf,He:2017spx,Geyer:2017ela,Geyer:2019hnn}, 
as it vastly reduces the number of independent
diagrams required in the computation \cite{Carrasco:2015iwa}. When applied
to perturbative gravity the double copy construction provides more
drastic boost in efficiency, as it allows alternative calculation
using gauge field kinematic numerators, thereby circumventing the
otherwise inconceivable infinite number of vertices present in the
amplitude computations. 

Despite its importance, a systematic construct of the kinematic numerator
for generic BCJ satisfying amplitude or the field theory all loop
level understanding of the duality remains to a large extent an open
question. At the time of writing loop level verification of the duality
among field theory amplitudes has acquired a great number of supporting
evidence through case by case explicit numerator solving with the
help of generalised unitarity techniques and in specific asymptotic
limit verified to all loop orders \cite{Saotome:2012vy,Oxburgh:2012zr}. One of the
puzzles lies at the heart of the understanding of the duality is to
find an explanation to the apparent Lie algebra-like feature presented
by kinematic numerators and has inspired numerous research from a
wide range of interesting perspectives 
\cite{Monteiro:2011pc,BjerrumBohr:2012mg,Tolotti:2013caa,Monteiro:2013rya,Bjerrum-Bohr:2016axv,Ho:2015bia,Fu:2016plh,Chen:2019ywi}.
An insight of great importance to the current paper is the string
theory explanation. The concept is to exploit the fact that string
amplitudes coincide with their field theory counterparts in the $\alpha'\rightarrow0$
point particle limit and BCJ amplitude identity (\ref{eq:bcj-amp-id})
can be seen as the corresponding limit of string amplitude monodromy
relations \cite{BjerrumBohr:2009rd,Stieberger:2009hq,BjerrumBohr:2010zs,Srisangyingcharoen:2020lhx}, which incidentally was known as
the Plahte identities \cite{Plahte:1970wy} in earlier literature.
Recent progress in this direction has extended the monodromy explanation
to arbitrary loop levels \cite{Tourkine:2016bak,Hohenegger:2017kqy,Ochirov:2017jby} and was shown by Casali,
Mizera and Tourkine \cite{Casali:2019ihm} that the monodromy relation
in turn can be naturally described by twisted homology $H_{m}(\mathcal{C}_{n,m}(z),\Phi_{\kappa})$
defined on a uni-valued branch of the Koba-Nielsen factor $\Phi_{\kappa}$
present in the string amplitude. 

\begin{figure}[t]
\centering
\subfigure[]{
\includegraphics[width=6cm]{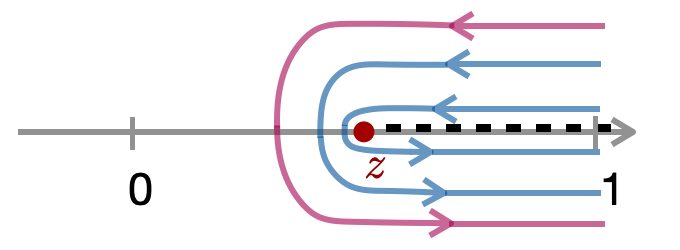}
\label{fig:multiple-screenings}   }
\subfigure[]{
\includegraphics[width=6cm]{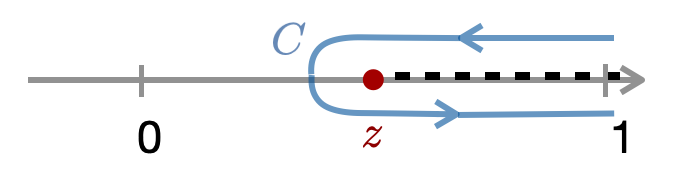} 
\label{fig:screening-def}   }
\caption{Integration contour of the screening.  }
\end{figure}

In a previous paper \cite{Fu:2018hpu} we extended the string theory
explanation one step further and suggested a natural string generalisation
of  kinematic numerator that plays a similar role in the analogue
of (\ref{eq:double-copy-1}) and presents the expected algebra-like
structure. The idea was to introduce a technique extensively used
to determine basis kinematic numerators (or \textit{master numerators})
in terms of partial amplitudes \cite{Du:2011js,Kiermaier-talk,Boels:2012sy,Du:2016tbc,Fu:2017uzt,Mafra:2015vca}. It is known that
once a pair of reference legs are chosen ($(1,n)$ in this case),
the set of $(n-2)!$ consecutive products of structure constants known
as the Del Duca-Dixon-Maltoni \cite{DelDuca:1999rs} half-ladders $f^{1,\sigma(2)}{}_{\rho}f^{\rho,\sigma(3)}{}_{\rho'}\dots f^{\rho'',\sigma(n-1)}{}_{n}$,
and therefore the set of kinematic numerators associated with the
same graphs, reproduces all $n$ point cubic tree graphs through repeatedly 
imposing Jacobi identities, where $\sigma(2),\sigma(3),\dots,\sigma(n-1)$
refers to the $S_{n-2}$ permutations of the $(n-2)$ non-reference
legs. Or formally speaking, there is a homomorphism from the set of
$n$ point BCJ numerators to the order $n-1$ Lie polynomials \cite{Frost:2019fjn}.The
half-ladder basis consists of the BCJ numerators whose images are
in the Lyndon-Shirshov basis under this homomorphism. Keeping this
in mind, note additionally that the $(n-2)!$ partial amplitudes $A(1,\sigma'(2),\sigma'(3),\dots,\sigma'(n-1),n)$,
being stripped of one copy of Chan-Paton factor, depends only on propagator
$D_{i}$'s and planar kinematic numerators with the specific ordering,
which in turn can be spanned by half-ladders. The rest of the partial
amplitudes can be solved using other amplitude identities known to
be compatible with the BCJ duality. The $(n-2)!$ by $(n-2)!$ coefficient
matrix that relates partial amplitudes and half-ladders is known to
coincide with the momentum kernel \cite{BjerrumBohr:2010ta,BjerrumBohr:2010yc} $\mathcal{S}[\sigma^{T}|\sigma']$
originally appeared in the gravity-gauge theory amplitude relation
discovered by Kawai, Lewellyn, and Tye (KLT) \cite{Kawai:1985xq}. As
a matter of fact from the BCJ perspective the KLT relation can be
understood as the double copy formulation of  a graviton amplitude
with its two copies of kinematic numerators being translated back
to gauge field partial amplitudes\footnote{A subtle technical issue actually arises because the coefficient propagator
matrix in the $(n-2)!$ half-ladder basis problem turns out to be
singular and prevent us from solving the numerators by direct inversion
\cite{Boels:2012sy}. In \cite{Fu:2018hpu} we bypassed this issue
by analytic continuing one of the legs and taking the invert while
the coefficient matrix in non-singular, at the cost of breaking the
$SL(2,\mathbb{R})$ invariance of the string amplitude temporarily.
The analytic continuation results in an overall $1/k_{n}^{2}$ factor
appears in the expression (\ref{eq:intro-half-ladd}). }. In light of the amplitude-numerator relation just described, in
\cite{Fu:2018hpu} we examine a natural generalisation of this relation
in the context of open string theory and wrote down the string analogue
of half-ladder numerators,
\begin{equation}
n(1,\sigma(2),\sigma(3),\dots,\sigma(n-1),n)=\lim_{k_{n}^{2}\rightarrow0}\frac{1}{k_{n}^{2}}\sum_{\sigma'\in S_{n-2}}\mathcal{S}_{\alpha'}[\sigma^{T}|\sigma']\,A(1,\sigma',n),\label{eq:intro-half-ladd}
\end{equation}
with the momentum kernel also replaced by its string analogue $\mathcal{S}_{\alpha'}[\sigma^{T}|\sigma']$
defined in \cite{BjerrumBohr:2010hn} as a product of sinusoidal factors
determined by the two sets of ordering $\sigma$ and $\sigma'$.
\begin{equation}
\mathcal{S}_{\alpha'}[\sigma_{1},\dots,\sigma_{k}|\sigma'_{1},\dots,\sigma'_{k}]:=\left(\frac{\pi\alpha'}{2}\right)^{-k}\prod_{t=1}^{k}\text{sin}\left(\pi\alpha'(k_{`}\cdot k_{\sigma_{t}}+\sum_{q>t}^{k}\theta(\sigma_{t},\sigma_{q})\,k_{\sigma_{t}}\cdot k_{\sigma_{q}})\right)
\end{equation}
Following similar argument used in \cite{BjerrumBohr:2010hn} we showed
that when multiplied with open string partial amplitudes, which are
Selberg integrals over ordered domains, the sinusoidal factors lift
contours to appropriate branches of the Riemann surface, so that the
half-ladder numerator (\ref{eq:intro-half-ladd}) is given by the
same integral as partial amplitude, but with the integration domain
combined into the multi-layer C-shaped contours (Fig.\ref{fig:multiple-screenings}),
which explains the expected half-ladder-like structure $f^{1,\sigma(2)}{}_{\rho}f^{\rho,\sigma(3)}{}_{\rho'}\dots f^{\rho'',\sigma(n-1)}{}_{n}$
when expressed as consecutive commutators of vertex operators integrated
over line contours along two sides of the branch cut.
\begin{equation}
\lim_{k_{n}^{2}\rightarrow0}\int_{0}^{1}\prod_{i=2}^{n-2}\frac{dz_{i}}{z_{i}}\Bigl\langle f\Bigr|[[[[V(z_{1}),V(z_{2})]_{\alpha'},V(z_{3})]_{\alpha'}\dots,V(z_{n-1})]_{\alpha'}\Bigl|0\Bigr\rangle\label{eq:qc}
\end{equation}
This consecutive quantum deformed commutator formula (\ref{eq:qc})
we arrived at suggests that BCJ numerators could be naturally derived
within the framework of an underlying quantum algebra. The existence
of such quantum algebraic structure is hinted by the twisted homological
representation of string amplitudes: As pointed out in \cite{Mizera:2017cqs,Mizera:2017rqa,Mizera:2019gea},
an n-point string amplitude can be viewed as (twisted) topological
invariants in the configuration space $\left\{ \left.\left(z_{2},..,z_{n-2}\right)\right|z_{i}\neq0,1,\quad z_{i}\neq z_{j}\right\} $
with a twist identical to the Koba-Nielson factor: 
Open string amplitudes are  pairings of one twisted cycle and one
twisted cocycle, and the closed string amplitudes are  pairings of
two twisted cocycles. Meanwhile, the isomorphism between twisted cycles
and quantum algebra modules is well known to the quantum algebra community 
(Detailed discussions can be found in monographs such as \cite{varchenko1995, Etingof:1998ru, varchenko:2003}). For example,  consider  a representation of the quantum algebra  $U_{q}\left(sl\left(2\right)\right)$  defined in (\ref{eq:apdx-ef-cr1}), (\ref{eq:apdx-ef-cr1}) and (\ref{eq:apdx-ef-cr3}), which is a tensor product of two Verma modules.  Such representation is generated by the tensor of the two highest weight vectors $v_{m}\otimes v_{l}$ . In this representation,  the vectors with weight $m+l-2$  
 are isomorphic to the twisted cycles in the
configuration $\mathbb{C}-\left\{ 0,1\right\} $ with twist factor
$z^{-m}\left(z-1\right)^{-l}$. 
(The notations used here will be explained in section \ref{sec:qa-review})
Under this isomorphism, $E v_{l}\otimes v_{m}$
and $v_{l}\otimes E v_{m}$ are mapped to the twisted cycle encircling
the branch points $0$ and $1$ respectively, and the boundary operator
on the twisted cycles is equivalent to the action of $F$ on the vectors.
This isomorphism between twisted homology and representation of quantum
algebra leads to the Drinfeld-Kohno theorem \cite{Drinfeld:1989st,Kohno87}: 
The solutions of Knizhnik-Zamolodchikov (KZ)
equations \cite{Knizhnik:1984nr}, as a representation of the homology group, is isomorphic
to the corresponding $R$-matrix representation of the quantum algebra.
This theorem is another hint for the quantum algebraic structure in
string amplitudes: At tree level, string amplitude takes the form
of generalised Selberg integrals, which appear as coefficients in
solutions for KZ equations \cite{varchenko1991}. Following the same spirit as Drinfeld-Kohno 
theorem, by matching string amplitudes with KZ coefficients,
we can read off the root system of the Lie algebra underlying KZ equations,
then the quantum deformation of this Lie algebra will naturally characterise
the behaviour of string amplitudes. 

In this paper we explore the relation between the basic concept of 
quantum algebra and string amplitudes 
by finding the root system for the kinematic algebra associated with
specific string amplitudes, showing
how the defining structure of quantum algebra is represented by 
contour integrals of the open string vertex operator, known as 
the screening operator in the quantum algebra literature 
\cite{Dotsenko:1984nm,Dotsenko:1984ad,Feigin:1981st}.
In addition we articulate how string amplitude
is related to the solution of KZ equations with the same root system.
We organise this paper as the following.
In section \ref{sec:screenings}, we begin with a minimal review of the quantum algebra,  
followed by 
an introduction of 
two slightly different versions of
the screening operators relevant to the discussion.
Then by identifying spacetime
momentum of the external legs as roots of the quantum algebra and deforming 
screening contours, we verified in section \ref{sec:bcj-numerators} that the screening
of the  vertex operators reproduce all defining
relations for representation of such quantum algebra, such as the
quantum deformed Lie brackets and the coproduct for the quantum universal
enveloping algebra.
In section \ref{sec:kz}, we discussed the relation
between Z-amplitude and the solutions to KZ equations with a brief
introduction to the KZ equation in 4.1 followed by the explicit $4$ point
and $5$ point examples . Finally we conclude our paper in section \ref{sec:conclusions}.

\section{Preliminaries}
\label{sec:screenings}
\subsection{Quantum algebra: A quick summary}
\label{sec:qa-review}

In this section we briefly review the notion of quantum groups. Historically
quantum groups first appeared in the inverse scattering problem of
integrable systems \cite{Faddeev:1979gh,Takhtajan:1979iv,Faddeev:1980zy} and was independently discovered
by Drinfeld \cite{Drinfeld:1986in} and Jimbo \cite{Jimbo:1985zk} when generalising
classical Lie algebra through deformations. Numerous equivalent formulations
are known \cite{Drinfeld:1986in,Jimbo:1985zk,Faddeev:1987ih,Manin1988,Manin:1989sz,Woronowicz:1987vs,Woronowicz:1987wr},
each emphasise on different aspects of the algebra. We adopt the approach
taken by Faddeev, Reshetikhin and Takhtajan (FRT) \cite{Faddeev:1987ih}
as it is formulated in a language that is more familiar to the readers
with physics background.

The modern definition of the quantum groups is a Hopf algebra $\mathcal{H}$
that satisfies additionally the quasi-triangular condition.
\begin{equation}
R\,\Delta(x)=\Delta'(x)\,R\quad,\text{ for all }x\in\mathcal{H},\label{eq:apdx-quasi-triang}
\end{equation}
where $\Delta'=\sigma\circ\Delta$ is the coproduct with the two factors of  its result
permuted, and $R\in\mathcal{H}\otimes\mathcal{H}$ is a tensor element
that plays the special role in the definition of the algebra called
the universal $R$-matrix. To motivate the definition, consider the
matrix elements of classical Lie group representation $T_{i\,j}(g)$,
regarded as complex functions of group element $g$. The polynomials
of $T_{i\,j}$'s is known to be a Hopf algebra, albeit a relatively
trivial one, with the coproduct, counit, antipode defined as
\begin{align}
\Delta(T_{i\,j})(g,g') & =T_{i\,k}(g)\otimes T_{k\,j}(g'),\label{eq:apdx-hopf-1}\\
\epsilon(T_{i\,j}) & =\delta_{i\,j},\label{eq:apdx-hopf-2}\\
S(T_{i\,j}) & =T_{i\,j}^{-1}.\label{eq:apdx-hopf-3}
\end{align}
The matrix elements are themselves $\mathbb{C}$-numbers, therefore
commutative, even though the matrix as a whole is not. The idea of
FRT is to generalise $T_{i\,j}$ to incorporate non-abelian algebra
(operators) by introducing a deformation through what is usually called
the $RTT$ relation,
\begin{equation}
R_{i\,j,\,k\,\ell}T_{k\,m}T_{\ell\,n}=T_{j\,\ell}T_{i\,k}R_{k\,\ell,\,m\,n}.\label{eq:apdx-rtt}
\end{equation}
For example in the simplest case where the universal $R$-matrix is
holomorphic in the indeterminate $h$, $R(h)=1+hR^{(1)}+h^{2}R^{(2)}+\dots$,
the $RTT$ relation (\ref{eq:apdx-rtt}) reads
\begin{equation}
T_{i\,m}T_{j\,n}-T_{j\,n}T_{i\,m}=h\cdot\left[R_{i\,j,\,k\,\ell}^{(1)}T_{k\,m}T_{\ell\,n}+T_{j\,\ell}T_{i\,k}R_{k\,\ell,\,m\,n}^{(1)}\right]+\mathcal{O}(h^{2}).
\end{equation}
Assuming the deformed $T_{i\,j}$ is also holomorphic, the above relation
implies its $h\rightarrow0$ limit is commutative, which is consistent
with the complex number representation of the classical Lie group.
The commutation relations of $T_{i\,j}$'s can be solved order by
order once an explicit $R$-matrix is chosen, starting with the classical
matrix element as its zeroth order $T_{i\,j}^{(0)}$. The $T_{i\,j}$'s
thus obtained was proved to satisfy the quasi-triangular condition
(\ref{eq:apdx-quasi-triang}) and at the same time preserves the Hopf
algebra structure (\ref{eq:apdx-hopf-1}), (\ref{eq:apdx-hopf-2}),
(\ref{eq:apdx-hopf-3}) at every order \cite{Faddeev:1987ih}. Note
from this perspective all classical Lie group algebras are actually
(trivial) quantum groups with the $R$-matrix being simply the identity.
Generically $T_{i\,j}$ can be a formal Laurent series in $h$, or
sometimes deformed through $q=e^{h}$, so that $T_{i\,j}\in\mathcal{H}[[q,q^{-1}]]$.
The Yangian $Y(g)$, in particular, familiar to the amplitude community
as the symmetry structure of the $\mathcal{N}=4$ super Yang-Mills
amplitude \cite{Drummond:2009fd} can be regarded as a special type
of quantum group defined by specific $R$-matrix.

In the literature very often the $RTT$ relation is abbreviated with
the help of introducing a convenient basis where $e^{i,j}\in End\,\mathbb{C}^{N}$
stands for the $N\times N$ complex matrix whose $i\,j$-th entry
is $1$ and otherwise zero. Denote
\begin{equation}
T_{a}:=\sum_{i,j}\underset{a-1}{\underbrace{1\otimes\dots\otimes1}}\otimes e^{i,j}\otimes\underset{m-a}{\underbrace{1\otimes\dots\otimes1}}\otimes T_{i\,j}
\end{equation}
as the algebra-coefficient matrix acting on $m$-tuple tensor of complex
vectors, $T_{a}\in\left(End\,\mathbb{C}^{N}\right)^{\otimes m}\otimes\mathcal{H}[[q,q^{-1}]]$,
where the coefficient is contracted with the $a$-th factor, the $RTT$
relation can be written in this language as
\begin{equation}
RT_{1}T_{2}=T_{2}T_{1}R.
\end{equation}
The same spirit is applied when describing matrices with tensor algebra
coefficients. For example $R_{0[13]}$, or more often written simply
as $R_{13}$ when no confusion can occur, refers to the $0$-tuple
matrix (namely a scalar, $1$) with its coefficient in $\mathcal{H}\otimes1\otimes\mathcal{H}$.
As consistency to the quasi-triangular condition (\ref{eq:apdx-quasi-triang}),
it is understood that the $R$-matrix needs satisfy the quantum Yang-Baxter
equation,
\begin{equation}
R_{12}R_{13}R_{23}=R_{23}R_{13}R_{12}.\label{eq:apdx-yb}
\end{equation}
Additionally, it is known that the center of the algebra is generated
by deformation parameter expansion of its quantum determinant $\text{det}_{q}\,T:=\sum_{\sigma\in S_{N}}(-q)^{\text{length}\sigma}T_{1\sigma(1)}\dots T_{n\sigma(N)}$,
which was elegantly used by Chicherin, Derkachov and Kirschner in
\cite{Chicherin:2013ora} to prove Yangian symmetry of the $\mathcal{N}=4$
super Yang-Mills amplitude recursively.

The quantum groups obtained from classical Lie group representation
through the deformation described above are called the \textit{quantum
matrix group}. The simplest example being the deformed $SL(2)$, or
$SL_{q}(2)$. In this case the $R$-matrix satisfying the Yang-Baxter
equation (\ref{eq:apdx-yb}) is given by $R_{i\,j,\,k\ell}=\delta_{i\,\ell}\delta_{j\,k}-q^{-2}\left(\begin{array}{cc}
 & 1\\
-q
\end{array}\right)_{j\,i}\left(\begin{array}{cc}
 & 1\\
-q
\end{array}\right)_{k\,\ell}$, and the $RTT$ relation (\ref{eq:apdx-rtt}) when read off component
by component gives
\begin{eqnarray}
ab=q\,ba, & \hspace{0.5cm} cd=q\,dc, & \hspace{0.5cm} ac=q\,ca,\label{eq:apdx-sl2}\\
bd=q\,db, & \hspace{0.5cm} bc=cb, & \hspace{0.5cm} ad=da+(q-q^{-1})bc,\nonumber
\end{eqnarray}
where $a$, $b$, $c$, $d$ are the entries of the $SL_{q}(2)$ matrix,
$T=\left(\begin{array}{cc}
a & b\\
c & d
\end{array}\right)$. 
The $R$-matrix corresponding to deforming $N\times N$ matrices
has been worked out so that $GL_{q}(N)$ can be derived similarly. 
(See \cite{Dobrev:2017qjn} for example.)

The quantum groups considered in this paper is actually the dual of
the quantum matrix group, $\mathcal{H}^{*}=\text{Hom}(\mathcal{H},\mathbb{C})$,
in the same spirit that Lie algebra is dual to the group manifold, and
with the multiplication, unit, coproduct, counit defined as the pullback
of the coproduct, counit, multiplication, unit respectively: For all
$x,y\in\mathcal{H}^{*},\,a,b\in\mathcal{H}$ we have
\begin{eqnarray}
\left\langle xy,a\right\rangle =\left\langle x\otimes y,\Delta(a)\right\rangle , &
\hspace{0.5cm} \left\langle I,a\right\rangle =\epsilon(a),\label{eq:apdx-frt1}\\
\left\langle \Delta(x),a\otimes b\right\rangle =\left\langle x,ab\right\rangle , &
\hspace{0.5cm} \epsilon(x)=\left\langle x,I\right\rangle ,\label{eq:apdx-frt2}\\
\left\langle S(x),a\right\rangle =\left\langle x,S(a)\right\rangle . & 
\hspace{0.5cm} \label{eq:apdx-frt3}
\end{eqnarray}
It follows from the properties of a Hopf algebra that its dual $\mathcal{H}^{*}$
is also a Hopf algebra. Conventionally $\mathcal{H}^{*}$ is arranged
into an upper and a lower triangular matrix $L_{i\,j}^{\pm}$ and
normalised with the help of the $R$-matrix (in the case of quantum
matrix groups happens to contain only $\mathbb{C}$-numbers)
\begin{equation}
\left\langle L_{i\,j}^{\pm},T_{k\,\ell}\right\rangle =R_{i\,k,\,j\,\ell}^{\pm},
\end{equation}
where $R_{i\,k,\,j\,\ell}^{+}:=q^{-1/2}R_{k\,i,\,\ell\,j}$ and $R_{i\,k,\,j\,\ell}^{-}:=q^{1/2}R_{i\,k,\,j\,\ell}^{-1}$.
It follows from the definition that $L_{i\,j}^{\pm}$'s also satisfy
$RTT$ relation (which in this case is sometimes referred as the $RLL$
relations)
\begin{align}
RL_{2}^{\pm}L_{1}^{\pm} & =L_{1}^{\pm}L_{2}^{\pm}R,\label{eq:apdx-rll-1}\\
RL_{2}^{+}L_{1}^{-} & =L_{1}^{-}L_{2}^{+}R,\label{eq:apdx-rll-2}
\end{align}
and therefore themselves is also a quantum group. For $SL_{q}(2)$
(\ref{eq:apdx-sl2}), its dual $L_{i\,j}^{\pm}$'s are conventionally
written as the following.
\begin{equation}
L^{+}=\left(\begin{array}{cc}
q^{\frac{-1}{2}H} & q^{\frac{-1}{2}}(q-q^{-1})X^{+}\\
0 & q^{\frac{1}{2}H}
\end{array}\right),\,L^{-}=\left(\begin{array}{cc}
q^{\frac{1}{2}H} & 0\\
q^{\frac{1}{2}}(q^{-1}-q)X^{-} & q^{\frac{-1}{2}H}
\end{array}\right),
\end{equation}
and their commutation relations are read off from the $RLL$ relations
(\ref{eq:apdx-rll-1}), (\ref{eq:apdx-rll-2}), similarly to the case
of quantum matrix group.
\begin{align}
[X^{+},X^{-}] & =[H]_{q},\label{eq:apdx-usl2-1}\\{}
[H,X^{\pm}] & =\pm2\,X^{\pm},\label{eq:apdx-usl2-2}
\end{align}
where in the above equations $[H]_{q}$ stands for the abbreviation
$[H]_{q}:=\frac{q^{H}-q^{-H}}{q-q^{-1}}$ called a $q$-number, as
the same combinations appear frequently in quantum group related calculations.
The similarity between the relation of the Lie group and Lie algebra
with that of their corresponding $q$-deforms extends actually beyond
an analogy. In the classical limit $q\rightarrow1$ (namely, $h\rightarrow0$)
equations (\ref{eq:apdx-usl2-1}) and (\ref{eq:apdx-usl2-2}) reduce
to the familiar commutation relations of the classical $sl(2)$ Lie
algebra. The associative algebra over $\mathbb{C}$ generated by $X^{\pm}$
and $H$ modulo (\ref{eq:apdx-usl2-1}) and (\ref{eq:apdx-usl2-2})
is called the quantum universal enveloping algebra (QUEA) $U_{q}(sl(2))$
for $sl(2)$, or \textit{quantum algebra} for shortness. Another related
formulation of the quantum algebra due to Jimbo \cite{Jimbo:1985zk} is
expressed in terms of the generators $E\sim X^{-}q^{H/2}$, $F\sim X^{+}q^{-H/2}$
and $K^{\pm}=q^{\pm H}$, properly normalised so as to satisfy the
following commutation relations.
\begin{align}
[E,F] & =\frac{K-K^{-1}}{q-q^{-1}},\label{eq:apdx-ef-cr1}\\
KEK^{-1} & =q^{-2}E,\label{eq:apdx-ef-cr2}\\
KFK^{-1} & =q^{2}\,F,\label{eq:apdx-ef-cr3}
\end{align}
which also coincide with the classical $sl(2)$ Lie algebra commutation
relations in the $q\rightarrow1$ limit. For our purposes we can simply
take the above as the definition of the QUEA as it is exactly the
starting point taken by Drinfeld and Jimbo \cite{Drinfeld:1986in,Jimbo:1985zk}.
For arbitrary algebra root systems the commutation relations are given
by (\ref{eq:screening-cr-1}), (\ref{eq:screening-cr-2}), (\ref{eq:screening-cr-3})
and (\ref{eq:screening-cr-4}). The Hopf algebra relations of the
QUEA can be worked out from the FRT prescription (\ref{eq:apdx-frt1}),
(\ref{eq:apdx-frt2}), (\ref{eq:apdx-frt3}). In the case of $U_{q}(sl(2))$
these are given by
\begin{eqnarray}
\Delta E=1\otimes E+E\otimes K, & \hspace{0.5cm} 
S(E)=-EK^{-1}, & \hspace{0.5cm}
\epsilon(E)=0, \label{eq:apdx-hopf1}\\
\Delta F=K^{-1}\otimes F+F\otimes1, & \hspace{0.5cm}
S(F)=-KF, & \hspace{0.5cm}
\epsilon(F)=0,\label{eq:apdx-hopf2}\\
\Delta K=K\otimes K, & \hspace{0.5cm}
S(K)=K^{-1}, & \hspace{0.5cm}
\epsilon(K)=1.\label{eq:apdx-hopf3}
\end{eqnarray}

\textit{Remark}. Note that the Hopf algebra structure naturally arises
from the symmetry algebra of multi-particle quantum system represented
as tensor of Hilbert space vectors $\Bigl|1\Bigr\rangle\otimes\Bigl|2\Bigr\rangle$
or complex vectors for spin states. A rotation generator for example,
satisfies $\Delta(J_{i})=J_{i}\otimes1+1\otimes J_{i}$, or very often
written as $J_{i}^{\text{total}}=J_{i}^{(1)}+J_{i}^{(2)}$ ($i=+,-,z$)
in physics literature. The defining conditions of the Hopf algebra
such as homomorphism of the coproduct ensures that when exponentiated,
generators produce the appropriate two particle rotation operator
$\Delta(e^{-iJ_{i}\theta})=e^{-iJ_{i}\theta}\otimes e^{-iJ_{i}\theta}$,
and that the cocommutative condition ensures a unique $n$-particle
operator $\Delta^{(n)}(e^{-iJ_{i}\theta})$ and so on. From this perspective
the quantum algebra, in particular its Hopf algebra relations (\ref{eq:apdx-hopf1}),
(\ref{eq:apdx-hopf2}), (\ref{eq:apdx-hopf3}), which also reduce
to their classical counterparts in the $q\rightarrow1$ limit, can
be seen as a non-trivial generalisation of the symmetry algebra of
multi-particle system. Indeed, as we should see in the following discussions
that the tensor of a string Hilbert state with an $SL(2,\mathbb{R})$
fixed vertex operator $V(z)$ (also known as an \textit{evaluation
module} \cite{Etingof:1998ru}) serves as a non-trivial two-particle
representation of the QUEA $\Bigl|1\Bigr\rangle\otimes\Bigl|2\Bigr\rangle$
with the deformation parameter given by braiding factor $q=e^{-i\pi\alpha'}$
produced by string monodromy, while the rest of the string insertions,
being integrated over $C$-shaped contours in the kinematic numerator,
act on the two particle state as symmetry algebra generators $E_{i}$.

In this paper we do not assume particular value for the inverse string
tension $\alpha'$, especially we rely on $\alpha'$ being able to
adjust freely to obtain field theory amplitudes from string theory
ones as their point particle limit, so that generically $q$ not being
a root of unity, in which case the representation of the quantum algebra
is known to be given by the Verma module $M_{\lambda}^{q}=\text{Ind}_{H\oplus F}\mathbb{C}_{\lambda}$
: Starting with a highest weight, namely highest eigenvalue, $1$-dimensional
complex vector $v_{\lambda}$,
\begin{equation}
H\,v_{\lambda}=\lambda\,v_{\lambda},\,F\,v_{\lambda}=0,
\end{equation}
the irreducible representation of the QUEA $U_{q}(sl(2))$ contains
all vectors freely generated by $E$: $E\,v_{\lambda}$, $E^{2}\,v_{\lambda}$,
$\dots$. We distinguish vectors by actions of $E$'s instead of labeling
vectors by eigenvalues like in the case of angular momentum algebra,
so that no confusion occurs when generalised beyond $sl(2)$ where
multiple non-mutually commuting $E_{i}$'s are present. The representation
generically is infinite dimensional and only becomes finite when $\lambda$
is an integer multiple of the deformation parameter $h$. 


\subsection{(i) C-shaped contour screening operators}
\label{sec:c-shaped-screenings}

Suppose if we denote the open string Hilbert space as $\mathfrak{H}$,
the vertex operators are algebra-valued distributions $\mathcal{V}\in End(\mathfrak{H})[[\tau,\tau^{-1}]]$
satisfying the state-field correspondence principle. For the purpose
of discussion let us for the moment focus on bosonic open strings,
so that a typical vertex operator takes the explicit form
\begin{equation}
V_{p}(z)=u\,e^{ip\cdot X(z)},\:u\in\{1,\,\epsilon\cdot\partial X,\,(\epsilon\cdot\partial X)^{2},\,\dots\},\label{eq:screening-module}
\end{equation}
with $z=e^{i\tau}$. A screening operator $E_{i}\in End(\mathcal{V})$ associated with
any vertex operator $S_{i}(t)=u\,e^{ik_{i}\cdot X(t)}$ of interest
is defined as the following integral over the C-shaped contour along
both sides of the branch cut \cite{Dotsenko:1984nm,Dotsenko:1984ad}.
\begin{align}
E_{i}\,V_{p}(z)\, & :=\int_{C}dt\,V_{p}(z)\,S_{i}(t).\label{eq:screening-op}\\
 & =\int_{C}dt\,(z-t)^{\alpha'p\cdot k_{i}}f(t)\,:V_{p}(z)S_{i}(t):\nonumber 
\end{align}
In this paper we follow the same conventions as in \cite{Green:1987sp} and
\cite{BjerrumBohr:2010hn}. Variables appear on the left are assumed
to start with a larger value on the real line than those on the right,
correspondingly shifted by a larger $i\epsilon$ on the complex plane,
and then analytically continued to their designated values. Note however
that in terms of figures, conventionally the real line points to the
right instead of left, so that everything illustrated in figures will
be the mirror image to what appears in the equation. For example the
action of $E_{i}$ is represented by Fig.\ref{fig:screening-def}. 
The analytic continuation we use here leads to the following
braiding relations for $z_{1}>z_{2}$.
\begin{align}
S_{i}(z_{1})\,S_{j}(z_{2}) & =e^{i\pi\alpha'k_{i}\cdot k_{j}}\,S_{j}(z_{2})\,S_{i}(z_{1})\label{eq:braiding-1}\\
S_{i}(z_{1})\,V_{p}(z_{2}) & =e^{i\pi\alpha'k_{i}\cdot p}\,V_{p}(z_{2})\,S_{i}(z_{1}).\label{eq:braiding-2}
\end{align}
In the presence of successive actions, the contour associated with the
operator that comes later is defined so as to surround the pre-existing
contours (Fig.\ref{fig:multiple-screenings}).

The action of an operator $F_{i}$ is defined to annihilate the contour
integral created by $E_{i}$ using the conformal property\footnote{
 The operator $L_{-1}$ used here refers to the Virasoro generator $L_{m}=\frac{1}{2}\sum_{-\infty}^{\infty}\alpha_{m-n}\cdot\alpha_{n}$  
 }
 of vertex
operator that reduces an integral to its boundaries, $[L_{-1},\int_{C}dt\,S_{i}(t)]=\int_{C}dt\,\partial S_{i}(t)$.
Explicitly this is defined to carry a normalisation factor so that
\begin{align}
S_{i}(1)F_{i}\left(\dots E_{i}\dots V_{p}(z)\right) & :=\frac{1}{e^{i\pi\alpha'k_{i}^{2}}-e^{-i\pi\alpha'k_{i}^{2}}}\left(V_{p}(z)\dots[L_{-1},\int_{C}dt\,S_{i}(t)]\dots\right)\label{eq:def-fi}\\
 & =\frac{e^{-i\pi\alpha'k_{i}\cdot p}-e^{i\pi\alpha'k_{i}\cdot p}}{e^{i\pi\alpha'k_{i}^{2}}-e^{-i\pi\alpha'k_{i}^{2}}}S_{i}(1)\left(\dots\dots V_{p}(z)\right).\nonumber 
\end{align}
The screenings thus defined together with the fixed point vertex operator
$V_{p}(z)$, which serves as the highest weight Verma module, provides
a representation of the QUEA
$U_{q}(g)$ \cite{Drinfeld:1983ky,Drinfeld:1985rx,Jimbo:1985zk,Jimbo:1985vd},
\begin{align}
E_{i}F_{j}-F_{j}E_{i} & =\delta_{i\,j}\,\frac{K_{i}-K_{i}^{-1}}{q_{i}-q_{i}^{-1}}\label{eq:screening-cr-1}\\
K_{i}E_{j} & =q_i^{-k_{i}\cdot k_{j}/k_i^2}E_{j}K_{i}\label{eq:screening-cr-2}\\
K_{i}F_{j} & =q_i^{k_{i}\cdot k_{j}/k_i^2}F_{j}K_{i}\label{eq:screening-cr-3}\\
K_{i}K_{j} & =K_{j}K_{i}\label{eq:screening-cr-4}
\end{align}
where $q_{i}:=e^{-i\pi\alpha'k_{i}^{2}}$and $K_{i}$ is the operator
that measures momentum, or \textit{charge} in the original settings
\cite{Feigin:1981st,Dotsenko:1984nm,Dotsenko:1984ad}, so that
$E_{i}$'s and $F_{i}$'s were supposed to lower or raise the background
charge produced by the module, hence the name screenings.
\begin{equation}
K_{i}:=exp-\oint\,k_{i}\cdot\partial X \label{eq:charge-def}
\end{equation}

\begin{figure}[t]
\centering
\includegraphics[width=12.5cm]{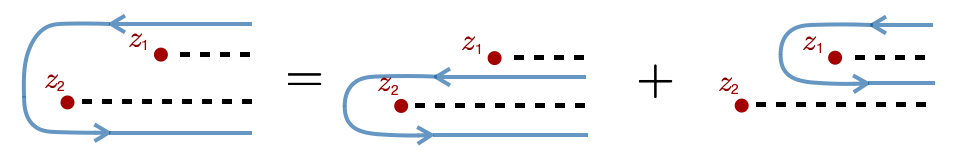}
\caption{Contour decomposition of the coproduct $\Delta E_{i}$. }
\label{fig:coproduct-defn}
\end{figure}

A natural representation for tensor of modules $v_{p}\otimes v_{s}$
can be obtained by simply taking the product of vertex operator at
distinct fixed points $V_{p}(z_{1})\,V_{s}(z_{2})$. The coproduct
of a screening $\Delta(E_{i})$ is then defined by the corresponding
action on this product followed by integration over a contour that
surrounds both vertices (Fig.\ref{fig:coproduct-defn}), which in
turn can be translated into the actions on individual modules by breaking
the original contour into two smaller ones surrounding each modules
and then swap the ordering using braiding relation (\ref{eq:braiding-2}).
\begin{align}
\Delta E_{i}\,\left(V_{p}(z_{1})\,V_{s}(z_{2})\right) & =\int_{C}dt\,V_{p}(z_{1})\,V_{s}(z_{2})\,S_{i}(t)\label{eq:coproduct-def}\\
 & =V_{p}(z_{1})\,\left[\int_{C}dt\,V_{s}(z_{2})\,S_{i}(t)\right]+e^{-i\pi k_{i}\cdot s}\left[\int_{C}dt\,V_{p}(z_{1})\,S_{i}(t)\right]\,V_{s}(z_{2}).\nonumber 
\end{align}
The above result is the same as the following tensors of screenings
\begin{equation}
\Delta E_{i}=1\otimes E_{i}+E_{i}\otimes K_{i},\label{eq:coproduct-e}
\end{equation}
as was expected for a quantum group. The action of antipode and counit
are represented by reversing and removing the contour of a screening
respectively.

\subsubsection{The $R$-matrix}

\label{sec:r-matrix}

\begin{figure}[t]
\centering
\includegraphics[width=9.3cm]{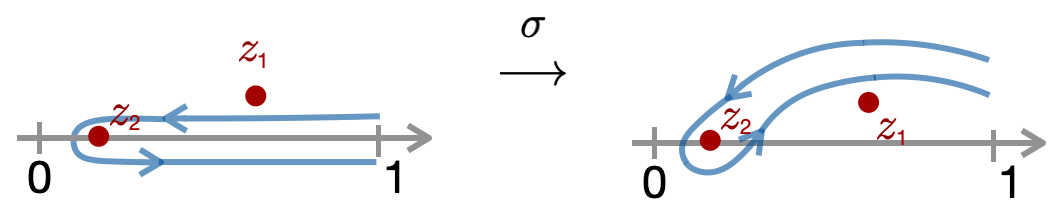}
\caption{Braiding effect of the $R$ matrix }
\label{fig:r-matrix}
\end{figure}

\begin{figure}[t]
\centering
\includegraphics[width=4.4cm]{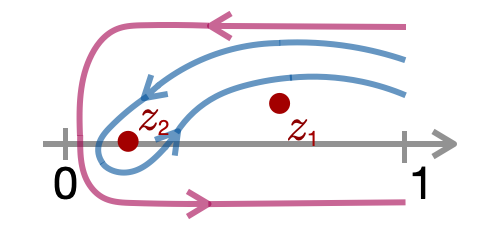}
\caption{Universal $R$-matrix acting on a coproduct}
\label{fig:quasi-triang}
\end{figure}

In the screening representation of quantum groups the universal $R$-matrix
is defined as the composition of a plain permutation $\sigma$ that
swaps modules along with their screenings, together with the application
of braiding relations (\ref{eq:braiding-1}) and (\ref{eq:braiding-2})
that eventually restores modules back to their original order.
\begin{equation}
v_{p}\otimes v_{s}\xrightarrow{\sigma}v_{s}\otimes v_{p}\xrightarrow{\mathfrak{R}}v_{p}\otimes v_{s}
\label{eq:r-matrix-mappings}
\end{equation}
Explicitly $\sigma$ maps for example  $V_{p}(z_{1})\left(E_{i}V_{s}(z_{2})\right)$
to $\left(E_{i}V_{s}(z_{2})\right)V_{p}(z_{1})$ as is illustrated
in Fig.\ref{fig:r-matrix}. The result can be re-expressed as integrals
over segments $t>z_{1}>z_{2}$ and $z_{1}>t>z_{2}$, which in turn
can be spanned by $V_{p}(z_{1})\left(E_{i}V_{s}(z_{2})\right)$ and
$\left(E_{i}V_{p}(z_{1})\right)V_{s}(z_{2})$ once braiding relations
were used. 
(The last procedure is denoted as $\mathfrak{R}$ in (\ref{eq:r-matrix-mappings}) so that the $R$-matrix is the composite $R=\mathfrak{R}\circ \sigma$.)
\begin{align}
V_{p}(z_{1})\left(E_{i}V_{s}(z_{2})\right)\longmapsto & e^{-i\pi\alpha'p\cdot k_{1}}(e^{i\pi\alpha's\cdot k_{1}}-e^{-i\pi\alpha's\cdot k_{1}})\int_{t>z_{1}>z_{2}}S_{i}(t)V_{p}(z_{1})V_{s}(z_{2})\\
 & +(e^{i\pi\alpha's\cdot k_{1}}-e^{-i\pi\alpha's\cdot k_{1}})\int_{z_{1}>t>z_{2}}V_{p}(z_{1})S_{i}(t)V_{s}(z_{2})\nonumber \\
= & V_{p}(z_{1})\left(E_{i}V_{s}(z_{2})\right)+(e^{-i\pi\alpha's\cdot k_{1}}-e^{i\pi\alpha's\cdot k_{1}})\left(E_{i}V_{p}(z_{1})\right)V_{s}(z_{2})\nonumber 
\end{align}
The above effect is the same as the action that successively removes
screenings from one of the modules within the tensor product and then
reapplies them onto the other\footnote{For simplicity we have neglected 
here the braiding factor produced by swapping modules, which will 
result in an overall $e^{\frac{H_{i}\otimes H_{i}}{2}}$ in the $R$-matrix}.
\begin{equation}
R \sim 1\otimes1+(q_{i}-q_{i}^{-1})E_{i}\otimes F_{i}+\dots
\end{equation}
Generically the complete formula for $R$ can be derived term by term
following similar reasoning \cite{Drinfeld:1986in,Kirillov:1991ec}. The quasi-triangular
condition $R\Delta(E_{i})=\Delta'(E_{i})R$ 
 (equation (\ref{eq:apdx-quasi-triang}))  
can be seen from the fact
that 
braiding two fixed vertices $V_{p}(z_{1})V_{s}(z_{2})$ does not alter the contour of a screening that encompasses them both  
 (Fig.\ref{fig:quasi-triang}). In particular that Yang-Baxter
equation $R_{12}R_{13}R_{23}=R_{23}R_{13}R_{12}$ is indeed satisfied
can be seen from the fact that $R$ derives from braiding. 

\subsection{(ii) Line interval screenings}
\label{sec:line-screenings}

An alternative version of the screening operator \cite{Semikhatov:2011ie}
that turns out to be also relevant to our BCJ problem is defined as
the line integral over a fixed interval along the real line, for example
over $[0,1]$,
\begin{equation}
E_{i}\,V_{p}(z)\,:=\int_{0}^{1}dt\,V_{p}(z)\,S_{i}(t),\label{eq:line-screening-def}
\end{equation}
whereas the charge operator is defined by the same closed integral
as before (\ref{eq:charge-def}). In the case of line interval screenings
it is sometimes convenient to restrict our considerations to only
the positive  Borel subalgebra $U_{q}^{+}(g)= H_{i}\oplus E_{i}$ 
of the full quantum universal enveloping algebra generated by Cartan
subalgebra and positive root vectors but without $F_{i}$'s, because
the boundary of a line interval is less symmetric than the C-shaped
contour, making it less natural to define the action of $F_{i}$ using
conformal generator as in (\ref{eq:def-fi}), even though one can
simply define it as manually removing one line screening. As we will
see in section \ref{sec:bcj-numerators} that the positive part of
the full algebra will be enough as far as numerators and amplitudes
are concerned. In these cases the screening operator
is to be interpreted as the insertion of an external particle
and there is no physical reason one must define an operator action that
removes particles. In the settings of line screenings tensors of modules
$v_{p}\otimes v_{s}$ are represented by products of vertex operators
$V_{p}(z_{1})\,V_{s}(z_{2})$ as before. Action of an $E_{i}$ on
individual module is defined similar to (\ref{eq:line-screening-def})
but with the integral carried out only to a manually fixed point $a$
between $z_{1}$ and $z_{2}$, whereas in the coproduct it is carried
out over the full segment $[0,1]$. The antipode and counit are represented
by reversing and removing the contour respectively as before. More
details  regarding line interval screenings 
can be found for example in \cite{Semikhatov:2011ie}.

\begin{figure}[t]
\centering
\includegraphics[width=6cm]{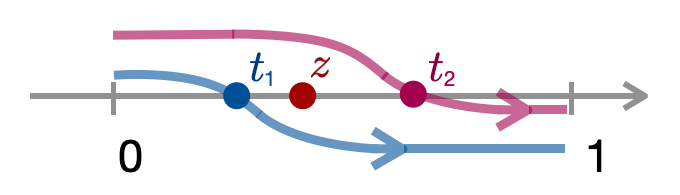}
\caption{The integration contours associated with the left and right actions
of screening. }
\label{fig:line-interval-screening-def}
\end{figure}

For the purpose of discussions it is useful to consider right action
of a line screening 
 $V_{p}(z)\,E_{i}$, defined the same as (\ref{eq:line-screening-def})
but with the ordering of the two vertex operators swapped. According
to our convention this corresponds to a line integral starting with
a point $t_{2}$ on the real line that is larger than $z$, and then
analytically continued, so that the contours associated with left
and right actions corresponds to the blue and lilac lines illustrated
in Fig.\ref{fig:line-interval-screening-def} respectively. From this
perspective a C-shaped screening can be identified as the $q$-deformed
adjoint action of line screening\footnote{ 
There are actually 
two different adjoint actions, $ad_{q}^{-}$ and $ad_{q}^{+}$,
that one can define for a quantum algebra, through taking the antipode of
the first and second tensor in (\ref{eq:adjoint-def}), respectively.
The two definitions correspond to the $q$ and $q^{-1}$-deformation of
the Lie bracket. In this paper the adjoint of interest is the $ad_{q}^{-}$
and we shall drop the minus sign to avoid excessive notation.  
},
\begin{equation}
ad_{q}(E_{i})=(m_{L}\otimes m_{R})\circ(S\otimes1)\circ\Delta(E_{i}),
\label{eq:adjoint-def}
\end{equation}
where $m_{L}$ and $m_{R}$ represent taking the left and right actions
respectively and $S$ is the antipode, $S(E_{i})=-E_{i}K_{i}^{-1}$.
Explicitly we have
\begin{align}
ad_{q}(E_{i})\,V_{p}(z) & =V_{p}(z)E_{i}-E_{i}K_{i}^{-1}V_{p}(z)K_{i}\nonumber \\
 & =(-1)e^{i\pi p\cdot k_{i}}\left[\int_{0}^{1}dt\,V_{p}(z)\,S_{i}(t)-e^{-i\pi p\cdot k_{i}}\int_{0}^{1}dt\,S_{i}(t)\,V_{p}(z)\right],
\end{align}
which is the same as (\ref{eq:screening-op}) up to an overall factor
that we will discard through redefinition.

\section{String BCJ numerators}

\label{sec:bcj-numerators}

In a previous paper \cite{Fu:2018hpu} we showed that the on-shell
limit of the multiple C-shaped contour integrals derived originally
from KLT in \cite{BjerrumBohr:2010hn} serves as a natural string theory
generalisation of the $(n-2)!$ basis BCJ
numerator.
\begin{equation}
n_{(n-2)!-\text{basis}}(1,2,3,\dots,n)=\int_{C_{i}}\prod_{i=2}^{n-2}dt_{i}\Bigl\langle f\Bigr|V_{1}(z_{1})\cdot V_{2}(t_{2})V_{3}(t_{3})\dots V_{n-1}(t_{n-1})\Bigl| 0 \Bigr\rangle , \label{eq:n-2-numerator-op}
\end{equation}
where $\Bigl\langle f\Bigr|$ is the asymptotic state at infinity, associated with leg $n$
through the state-field correspondence principle $\Bigl\langle f\Bigr|=\lim_{z\rightarrow\infty}\Bigl\langle0\Bigr|V_{n}(z)\,z$.
In the equation above we have slightly abused the notation. 
To every vertex operator $V_{i}(z_{i})$ we assumed that a specific string excitation mode 
(not necessarily tachyon) is implicitly assigned along with the momentum $k_{i}$. 
Additionaly  let us also recall that
the $(n-3)!$ field theory
basis numerator
was identified in the context of  Cachazo-He-Yuan (CHY) formulation of amplitudes 
in \cite{Cachazo:2013iea} as the product
of momentum kernel with partial amplitudes\footnote{  
The $\gamma$ on the right 
is understood to contain only the first $(n-3)$ indices,
$\gamma^{T}={\gamma(n-2),\dots,\gamma(2)}$ so that when all permutations are considered,
the momentum kernel $\mathcal{S}[\gamma^{T}|\beta]$  is an $(n-3)! \times (n-3)!$ matrix. 
 },
\begin{equation}
n(1,\gamma(2),\gamma(3),\dots,\gamma(n-1),n)=\begin{cases}
\sum_{\beta\in S_{n-3}}\mathcal{S}[\gamma^{T}|\beta]\,\tilde{\mathcal{A}}_{n}((1,\beta,n,n-1) & ,\:\gamma(n-1)=n-1\\
0 & ,\:\gamma(n-1)\neq n-1
\end{cases}\label{eq:n-3-numerator}
\end{equation}
When applied to strings, with both the momentum kernel and amplitudes substituted by
their string theory generalisations, the $(n-3)!$ basis numerator 
can be expressed also as multiple
C-shaped contour integrals \cite{BjerrumBohr:2010hn}, but with three
of the vertices $(z_{1},z_{2},z_{n})$ fixed instead of two\footnote{For the purpose of illustration we used here the antisymmetry of the
numerator and reversed its ordering, $n(1,2,,\dots,n-1,n)=(-1)^{n-2}n(n,n-1,\dots,2,1)$.},
\begin{equation}
n_{(n-3)!-\text{basis}}(1,2,3,\dots,n) \sim \int_{C_{i}}\prod_{i=2}^{n-2}dt_{i}\Bigl\langle 0 \Bigr|V_{n}(z_{n})\cdot V_{2}(z_{2})V_{1}(z_{1})\cdot V_{3}(t_{3})\dots V_{n-1}(t_{n-1})\Bigl| 0 \Bigr\rangle.\label{eq:n-3-numerator-op}
\end{equation}
For a specific string theory, the integrands in  (\ref{eq:n-2-numerator-op})
and (\ref{eq:n-3-numerator-op}) can be computed  by normal ordering
 vertex operators. The explicit formula for such integrands is the product of Koba-Nielsen factor $\prod (z_i-z_j)^{\alpha ' k_i\cdot k_j}$ and a linear combination rational functions composed of factors of the form $\frac{\epsilon_i\cdot k_j}{z_i-z_j}, \, \frac{\epsilon_i\cdot \epsilon_j}{(z_i-z_j)^2}$.
Generically the result of integration will be a linear combination of hypergeometric functions of $k_i\cdot k_j$ dressed with rational factors depending on $\epsilon_i\cdot k_j$ and $\epsilon_i\cdot\epsilon_j$. 
Suppose if we focus on the vertex operators in (\ref{eq:n-2-numerator-op})
and (\ref{eq:n-3-numerator-op}), ignoring for the moment a common
final leg $V_{n}(z_{n})$ that is frequently pushed to infinity, the
multiple C-shaped contour integrals appear in $(n-2)!$ and $(n-3)!$
basis numerators can be identified as screenings acting on single
and tensor modules respectively. Explicitly, for the $(n-3)!$ basis
numerator this is
\begin{equation}
\Delta(E_{n-1})\dots\Delta(E_{4})\Delta(E_{3})v_{k_{2}}\otimes v_{k_{1}}.\label{eq:screeningmodule}
\end{equation}
The above settings naturally defines a representation of quantum group
$U_{q}(g)$ with the $E_{3}$, $E_{4}$, $\dots$, $E_{n-1}$ identified
as the simple root vectors. Comparing with the definition of a screening
(\ref{eq:screening-op}) and equations (\ref{eq:screening-cr-1})
to (\ref{eq:screening-cr-3}) we see that the corresponding simple
roots are identified with the momenta $k_{3}$, $k_{4}$, $\dots$,
$k_{n-1}$ carried by external legs. In the infinite string tension
limit $\alpha'\rightarrow0$, therefore $q_{i}\rightarrow1$ and the
QUEA $U_{q}(g)$ reduces to the classical Lie (or Kac-Moody) algebra
$g$ with the (symmetrised) Cartan matrix defined by the same roots,
\begin{align}
[e_{i},\,f_{j}] & =\delta_{i\,j}\,h_{j}\label{eq:lie-cr-1}\\{}
[h_{i},\,e_{j}] & =(k_{i}\cdot k_{j})\,e_{j}\label{eq:lie-cr-2}\\{}
[h_{i},\,f_{j}] & =-(k_{i}\cdot k_{j})\,f_{j}\label{eq:lie-cr-3}
\end{align}
so that in the field theory limit the BCJ kinematic algebra should
be isomorphic to the algebra determined by external leg momenta. Note that we have used lower case letters for generators of classical Lie (Kac-Moody) algebra, and will keep using this convention in later discussions involving classical  Lie (Kac-Moody) algebra.

\begin{figure}[t]
\centering
\includegraphics[width=13.3cm]{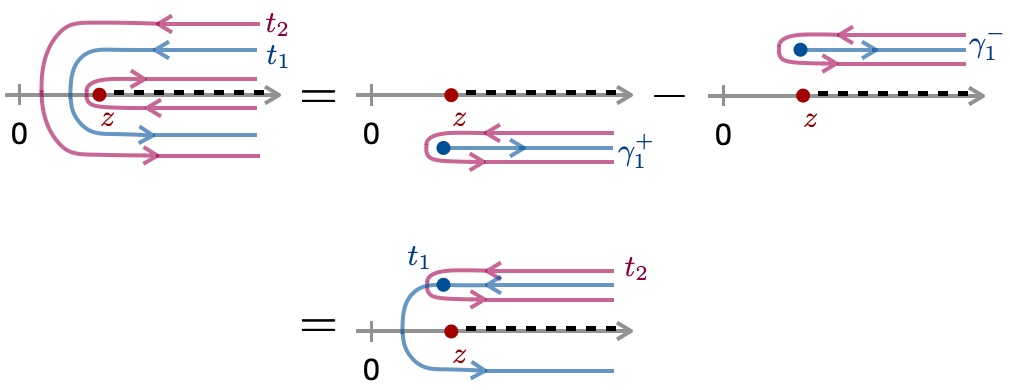}
\caption{Contours associated with the adjoint action}
\label{fig:adjoint-contour-2}
\end{figure}

Starting with $E_{3}$, $E_{4}$, $\dots$, $E_{n-1}$ as building
blocks the QUEA thus defined contains non-simple root vectors generated
by $q$-commutators $E_{k_{1}+k_{2}}\sim[E_{2},E_{1}]_{q}=ad_{q}(E_{2})E_{1}$.
These are root vectors in the sense that they satisfy similar commutation
relations (\ref{eq:screening-cr-2}), (\ref{eq:screening-cr-3}) and
$[E_{\gamma},\,F_{\gamma}]=C_{\gamma}\,[H_{\gamma}]_{q}$ up to a
normlisation  that depends on its root $\gamma$. 
(Here $[H_{\gamma}]_{q}$ stands for the $q$-number (\ref{eq:apdx-usl2-1})
defined in section \ref{sec:qa-review}.) 
The result of the
$q$-commutator can be seen from Fig.\ref{fig:adjoint-contour-2}
to be by itself a screening operator, but with its vertex operator
calculated from the following operator product.
\begin{align}
 & \int_{C_{2}\text{ encompassing }t_{1}}dt_{2}\,e^{ik_{1}\cdot X(t_{1})}e^{ik_{2}\cdot X(t_{2})}\nonumber \\
 & =sin\,\pi\alpha'k_{1}\cdot k_{2}\::e^{i(k_{1}+k_{2})\cdot X(t_{1})}\left(\frac{1}{k_{1}\cdot k_{2}+1}+\frac{k_{2}\cdot\partial X(t_{1})}{k_{1}\cdot k_{2}+2}+\frac{\frac{1}{2!}k_{2}\cdot\partial^{2}X(t_{1})}{k_{1}\cdot k_{2}+3}+\dots\right):\label{eq:alpha-beta-ope}
\end{align}
where we have chosen$E_{1}$, $E_{2}$ to be tachyons as an example.
This process continues generating new root vectors indefinitely until
it is terminated by quantum Serre relations $\left(ad_{q}(E_{i})\right)^{1-(k_{i}\cdot k_{j})}E_{j}=0$,
which in turn are determined by roots. For example the result of two
consecutive adjoint actions $ad_{q}(E_{3})\,ad_{q}(E_{2})\,E_{1}$
is to replace the vertex operator with the following.
\begin{align}
 & \int_{C_{2},C_{3}}dt_{2}dt_{3}\,S_{1}(t_{1})S_{2}(t_{2})S_{3}(t_{3})\nonumber \\
 & =(2i)^{2}\text{sin}\pi\alpha'k_{1}\cdot k_{2}\,\left[\text{sin}\pi\alpha'k_{1}\cdot k_{3}\,I(k_{1},k_{3},k_{2})+\text{sin}\pi\alpha'(k_{1}+k_{2})\cdot k_{3}\,I(k_{1},k_{2},k_{3})\right],\label{eq:sl3-serre}
\end{align}
where $I$'s are ordered operator product line integrals, 
\begin{equation}
I(k_{1},k_{2},k_{3})=\int_{t_{1}<t_{2}<t_{3}}(t_{3}-t_{2})^{k_{3}\cdot k_{2}}(t_{3}-t_{1})^{k_{3}\cdot k_{1}}(t_{2}-t_{1})^{k_{2}\cdot k_{1}}:S_{1}(t_{1})S_{2}(t_{2})S_{3}(t_{3}):
\end{equation}
and likewise for the other ordering. When the 
 momenta $k_{i}$'s are identified with for example, the simple roots $\alpha$ and $\beta$ 
of the $U_{q}(sl(3))$, namely if we choose $k_{1}=\beta$ and 
$k_{2}=k_{3}=\alpha$, 
we see that equation (\ref{eq:sl3-serre}) becomes
\begin{equation}
ad_{q}(E_{\alpha})^{2}E_{\beta}\sim\text{sin}\pi\alpha'(\beta\cdot\alpha)\,I(\beta,\alpha,\alpha)+\text{sin}\pi\alpha'((\beta+\alpha)\cdot\alpha)\,I(\beta,\alpha,\alpha)
\end{equation}
which indeed vanishes because for $U_q(sl(3))$ the roots satisfy $\alpha\cdot(\alpha+\beta)+\alpha\cdot\beta=0$,
and therefore $ \text{sin} \pi\alpha'((\alpha+\beta)\cdot\alpha)=-\text{sin}\pi\alpha'(\alpha\cdot\beta$).
For generic momentum configuration 
 there is no such identity to stop new non-simple roots
being generated and
the algebra $g$ is therefore 
infinite dimensional.
To make the algebra finite one can chose to work with compactified
spacetime such that all momenta live on a rational lattice \cite{Lentner:2017dkg},
in this case all $\alpha'k_{i}\cdot k_{j}$ eventually become integers
after being superposed  large enough number of times and the overall sinusoidal
factor appears in (\ref{eq:alpha-beta-ope}) vanishes. Note that the
BCJ amplitude relation can be regarded as a special type of the Serre
relations (\ref{eq:sl3-serre}) even though it eliminates only the
root vector that carries zero root constrained by momentum conservation
$k_{1}+\dots+k_{n}=0$, especially that roots containing multiple
copies of the same momentum $n_{1}k_{1}+n_{2}k_{2}+\dots$ remain
and the algebra is infinite unless rest of the constraints just described
were imposed.

In the settings of string theory asymptotic states provides a natural
definition for a bilinear form 
that can be used to normalise modules and root vectors.
For example when both states are gluons, from state-field correspondence 
we have
\begin{eqnarray}
(v_{p},v_{s}) & :=  & \lim_{z_{1}\rightarrow0,\,z_{n}\rightarrow\infty}z_{n}\Bigl\langle V_{-p}(z_{n})V_{s}(z_{1})\Bigr\rangle\frac{1}{z_{1}} \\
 & = & \Bigl\langle0\Bigr|\epsilon_{n}\cdot\alpha_{1}e^{-ip\cdot x}\,\epsilon_{1}\cdot\alpha_{-1}e^{is\cdot x}\Bigl|0\Bigr\rangle \nonumber
\end{eqnarray}
Note in particular
from this perspective the $(n-3)!$ basis numerator can be regarded
as a quantum Clebsch-Gordan coefficient. Different choices of vertex
operator screenings in this setting correspond to representations
associated with different modes. 
Generically
the bilinear form can be similarly defined for arbitrary values of
$z_{1}$ and $z_{n}$ and different choices of bilinear form are related
by KZ equations. We leave this part of the
discussion to section \ref{sec:correlators}.

\begin{figure}[t]
\centering
\includegraphics[width=8.4cm]{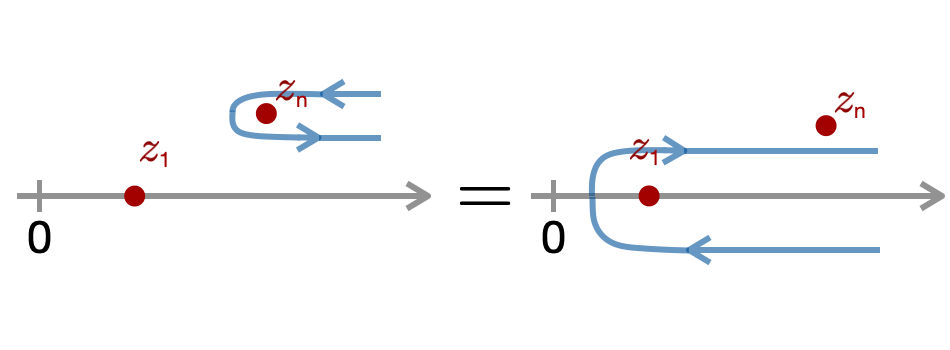}
\caption{Flipping the contour surrounding $z_{n}$.}
\label{fig:bilinear-form}
\end{figure}

\textit{Remark}. Before we proceed, we would like to clarify a subtle issue
related to the bilinear form defined above. The Hilbert space and
its dual inherited from string settings are Verma modules of the string
modes, which by themselves comprised a Heisenberg algebra. From this
perspective the screening operators are actually a representation
of the quantum algebra built on top of the Heisenberg algebra 
(referred to as the \textit{basic representation} in \cite{Frenkel:1980rn}).
While the string Hilbert space and its dual together defines a involutive
(Hermitian in this case) bilinear form $\left\langle \alpha_{-n}^{\mu}v_{p},\,v_{s}\right\rangle =\left\langle v_{p},\,\alpha_{n}^{\mu}v_{s}\right\rangle $,
the involution feature generically does not pass on to the screening
operators. The action of a quantum algebra 
root vector in this bilinear
form is not the same as the annihilation of the same root vector on
its dual space. Namely, we do not have the identity $\left\langle E_{k}v_{p},\,v_{s}\right\rangle =\left\langle v_{p},\,F_{k}v_{s}\right\rangle ,$
so that the bilinear form (and therefore the kinematic numerator)
is not directly computed from simple algebraic manipulations as angular
momentum algebra. As a matter of fact 
when the C-shaped contour extends to infinity
the action of a quantum root vector actually equals its antipode acting on the dual module,
which can be seen by flipping the contour surrounding one module to the other (Fig.\ref{fig:bilinear-form}). 
\begin{eqnarray}
(E_{k}v_{p},v_{s}) & =& -\int_{\text{surrounding }z_{1}}dt_{1}\,z_{n}\Bigl\langle0\Bigr|V_{p}(z_{n})S_{k}(t_{1})V_{s}(z_{1})\Bigl|0\Bigr\rangle\frac{1}{z_{1}}\\
 & = &(v_{p},-E_{k}K^{-1}v_{s}) \nonumber 
\end{eqnarray}
(An extra phase factor $K^{-1}$ was due to swapping $S_{k}(t_{1})$ and $V_{s}(z_{1})$)


\subsection{Jacobi-like identities}

In light of the original idea of BCJ duality it is perhaps more or
less expected that the numerator is expressible as successive adjoint
actions that mimics the colour dependence of the amplitude, and indeed
it was realised in \cite{Carrasco:2016ygv,Ma:2011um,Fu:2018hpu} that such structure is accounted
for in string theory by deformed brackets. In addition we note that
because the C-shaped screening can be regarded as the $q$-deformed
adjoint action of line screenings, the $(n-2)!$ basis numerator (\ref{eq:n-2-numerator-op})
can be recast into the following BCJ manifest form.
\begin{equation}
n_{(n-2)!-\text{basis}}(1,2,3,\dots,n)=\bigl\langle f\bigr|ad_{q}(E_{n-1})\dots ad_{q}(E_{3})\,ad_{q}(E_{2})\,V_{1}(t_{1})\bigl|0\bigr\rangle,
\end{equation}
where the $E_{i}$'s above are line screenings, and the numerator
is therefore the successive adjoint actions of QUEA determined by
external leg momenta. If our only purpose is to explain the BCJ duality
originally observed in field theory amplitudes, it is not strictly
necessary to come up with a $q$-deformed analogue of the Jacobi-like
identity satisfied by string numerators. However we do actually have
an identity that explains the expected relation at quantum level,
\begin{align}
ad_{q}(E_{i})\,ad_{q}(E_{j})\circ E_{\ell}-q^{k_{i}\cdot k_{j}}\,ad_{q}(E_{j})\,ad_{q}(E_{i})\circ E_{\ell} & =ad_{q}([E_{i},\,E_{j}]_{q})\circ E_{\ell}.\label{eq:jacobi-like-eq}
\end{align}
Surprisingly the above identity does not directly come from the perhaps
seemingly more natural candidate implied by the definition of $q$-commutator,
$[E_{a},\,[E_{b},\,E_{c}]_{q^{b\cdot c}}]_{q^{a\cdot c}}-q^{a\cdot b}[E_{b},\,[E_{a},\,E_{c}]_{q^{a\cdot c}}]_{q^{b\cdot c}}=[[E_{a},\,E_{b}]_{q^{a\cdot b}},\,E_{c}]_{q^{(a+b)\cdot c}}$,
as careful inspection would quickly show that the $q$-deformed factors
mismatch, but rather is the consequence of braiding relations (\ref{eq:braiding-1})
and (\ref{eq:braiding-2}). The Jacobi-like identity (\ref{eq:jacobi-like-eq})
can be verified by explicit translating screenings into ordered line
integrals and cancel.


\section{Relation to the KZ equations}
\label{sec:kz}

Recall that the KZ equations \cite{Knizhnik:1984nr}  is a set of differential equations for
Lie algebra (or more generally Kac-Moody algebra)-module $\phi$ with
coordinate dependence,
\begin{equation}
\frac{\partial\phi}{\partial z_{i}}=\frac{1}{\kappa}\sum_{j\neq i}\frac{\Omega_{ij}}{z_{i}-z_{j}}\phi.\label{eq:KZ equation}
\end{equation}
For a Kac-Moody algebra with simple roots $\left\{ k_{i}\right\} $
and Cartan subalgebra $\left\{ h_{i}\right\} $, satisfying the following
(classical) commutation relations,
\begin{align}
[h_{i},e_{k_j}]=-(k_{j})_{i}\,e_{k_j}, & \hspace{1cm} [h_{i},f_{k_j}]=(k_{j})_{i}\,f_{k_j},\nonumber\\{}
[e_{k_i},f_{k_j}]=\delta_{i\,j}(k_{j})_{i}\,h_{i}, & \hspace{1cm} [h_{i},h_{j}]=0.\label{eq:kac-moody-cr}
\end{align}
Its Verma module $\tilde{V}_{\Lambda}$ is generated by the lowest
weight module $\tilde{v}_{\Lambda}$, and root vectors $f_{k_i}$,
\begin{equation}
\tilde{v}_{nk_{i}+\Lambda}=(f_{k_i})^{n}\tilde{v}_{\Lambda}.
\end{equation}
The module $\phi$ that appears in the KZ equation is a map from the
space $U=\{(z_{1},z_{2},\dots,z_{n})\in\mathbb{C}^{n}|z_{i}\neq z_{j}\}$
to the space of tensor modules $\tilde{V}_{\Lambda_{1}}\otimes\tilde{V}_{\Lambda_{2}}\otimes\dots\otimes\tilde{V}_{\Lambda_{n}}$.
For example when $n=2$,
\begin{align}
\phi_{0} & =I_{0}\,\tilde{v}_{p_{1}}\otimes\tilde{v}_{p_{2}},\label{eq:top}\\
\phi_{1} & =I_{(1,0)}\,f_{k}\tilde{v}_{p_{1}}\otimes\tilde{v}_{p_{2}}+I_{(0,1)}\,\tilde{v}_{p_{1}}\otimes f_{k}\tilde{v}_{p_{2}} , \label{eq:first}\\
 & \vdots\nonumber
\end{align}
On the other hand the operator $\Omega$ in the KZ is the Casimir,
\begin{equation}
\Omega=\sum_{h_i\in\text{ Cartan subalgebra}}h_{i}\otimes h_{i}+\sum_{k_i\in\text{ all roots}}\left(f_{k_i}\otimes e_{k_i}+e_{k_i}\otimes f_{k_i}\right) .\label{eq:casimir}
\end{equation}
Note that  the second summation runs over all root vectors including non-simple ones, and $\Omega_{ij}$ is understood to act only on the $i$-th and $j$-th
factor of the tensor. By construction $\Omega_{ij}$ commutes with all
coproducts in the algebra, in particular $[\Omega,\Delta(e_{i})]=0$,
so that it only mixes tensors with the same overall weights. In light
of this the solutions $\phi$ can be assorted into ground modules, $1$-level
raised modules and so on, as was shown by equations (\ref{eq:top})
and (\ref{eq:first}). Explicitly the coefficient functions are given
by
\begin{equation}
I_{0}=(z_{2}-z_{1})^{p_{1}\cdot p_{2}},\,I_{(1,0)}=\int_{\gamma}dt\frac{1}{t-z_{1}}\Phi_{\kappa},\,I_{(0,1)}=\int_{\gamma}dt\frac{1}{z_{2}-t}\Phi_{\kappa},\label{eq:i-integrals}
\end{equation}
where $\Phi_{\kappa}=\left(t-z_{1}\right)^{k\cdot p_{1}/\kappa}\left(z_{2}-t\right)^{k\cdot p_{2}/\kappa}\left(z_{2}-z_{1}\right)^{p_{1}\cdot p_{2}/\kappa}$
and $\gamma$ is any closed contour. Generically the solution corresponding
to an $m$-level lowered module is given by integrals of $m$-forms
$\Phi_{\kappa}A_{m}dt_{i_{1}}\wedge dt_{i_{2}}\dots\wedge dt_{i_{n}}$
over a loop $\gamma$ in the punctured space.

\begin{figure}[t]
\centering
\includegraphics[width=10cm]{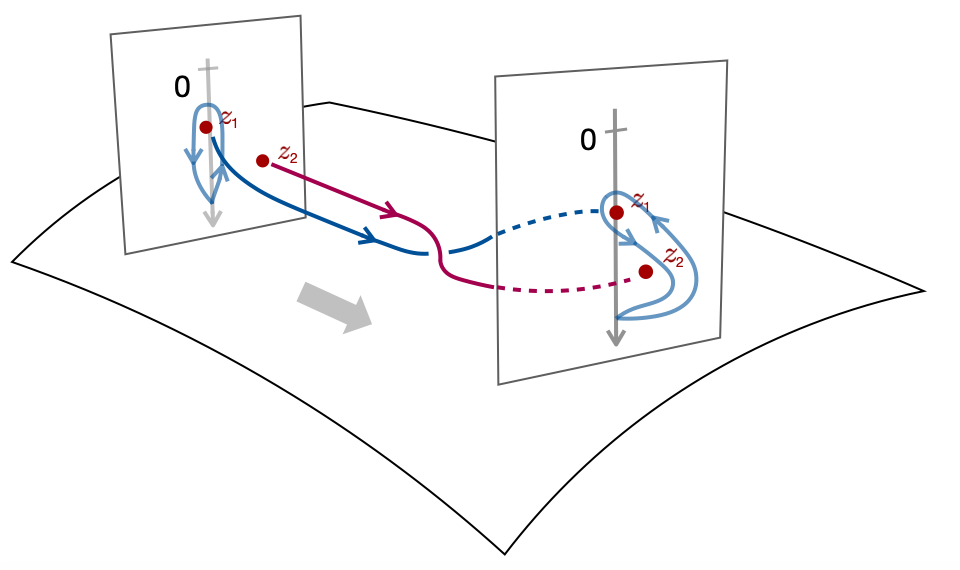} 
\caption{A braiding defined by the Knizhnik-Zamolodchikov connection}
\label{fig:bundle}
\end{figure}

The settings of KZ has a natural geometry interpretation, where $\phi$
can be identified as the horizontal section, $d\phi-\Gamma\phi=0$,
determined by the KZ flat connection $\Gamma=\sum_{i,j}\frac{\Omega_{ij}}{z_{i}-z_{j}}(dz_{i}-dz_{j})$.
Indeed if we consider a bundle with base $U$ and $\tilde{V}_{\Lambda_{1}}\otimes\tilde{V}_{\Lambda_{2}}\otimes\dots\otimes\tilde{V}_{\Lambda_{n}}$
as the fibre, starting with $\phi(z_{1}^{0},z_{2}^{0},\dots,z_{n}^{0})$
at a specific value of $z_{i}$'s, the solution to KZ equations
at generic point $\phi(z'_{1},z'_{2},\dots,z'_{n})$ can be obtained
through a unique lift. The coefficients $I_{i}$'s on the other hand,
defines a pairing between the twisted cohomology group $H^{m}(\mathcal{C}_{n,m}(z),\Phi_{\kappa})$
and homology group $H_{m}(\mathcal{C}_{n,m}(z),\Phi_{\kappa})$ on
the punctured space\footnote{The integration variables $t_{i}$'s appear in (\ref{eq:i-integrals})
and their higher level generalisations live in the space of discriminantal
arrangement $\mathcal{C}_{n,m}(z)$ \cite{varchenko:2003}.
When there is only one variable this space is simply the punctured
space $\mathbb{C}-\{z_{1},z_{2},\dots,z_{n}\}$. In the cases of multiple
$t_{i}$'s the integral generically would contain $(t_{i}-t_{j})^{k_{i}\cdot k_{j}}$
and we must impose additionally that $t_{i}\neq t_{j}$.} $\mathcal{C}_{n,m}(z)$, identified as the $m$-form and the closed
contour $\gamma$ respectively. An $I_{i}$ depends only on the (twisted)
homology once we have chosen a particular $m$-form, so that when
$z_{i}$'s vary along a path in the base space the contour $\gamma$
deforms continuously, and the action of the KZ provides a Gauss-Manin
connection on the bundle with base $U$ and twisted homology group $H_{m}(\mathcal{C}_{n,m}(z),\Phi_{\kappa})$
as its fibre. In particular when the end point $\{z_{1},z_{2},\dots,z_{n}\}$
is a permutation of the starting point $(z_{1}^{0},z_{2}^{0},\dots,z_{n}^{0})$,
the action of KZ braids $\gamma$ (Fig.\ref{fig:bundle}) and defines
an $R$-matrix on $H_{m}(\mathcal{C}_{n,m}(z),\Phi_{\kappa})$. The
twisted homology group is known to be isomorphic to the quantum group
$U_{q}(g)$ that corresponds to the $q$-deformation of (\ref{eq:kac-moody-cr}) \cite{varchenko1991}.

\subsection{Correlators, bilinear forms and KZ solutions}

\label{sec:correlators}

In this section we temporarily remove all integrals present in an
amplitude or BCJ numerators (\ref{eq:n-2-numerator-op}), (\ref{eq:n-3-numerator-op})
and focus exclusively on their correlator $\Bigl\langle 0 \Bigr|V_{p_{n}}(z_{n})\dots V_{p_{2}}(z_{2})V_{p_{1}}(z_{1})\Bigl| 0 \Bigr\rangle$.
When all vertices are tachyons apparently the correlator is the $0$-form
coefficient $I_{0}=\prod_{i,j}(z_{i}-z_{j})^{\alpha'k_{i}\cdot k_{j}}$
of the top highest weight module $\phi_{0}=I_{0}\,v_{p_{n}}\otimes\dots\otimes v_{p_{2}}\otimes v_{p_{1}}$
and the KZ equations (\ref{eq:KZ equation}) in this case translate
to the differential equations of $I_{0}$, whereas the action of Casimir
$\Omega_{ij}$ can be read off directly from the module, giving
\begin{equation}
\frac{\partial}{\partial z_{i}}I_{0}=\frac{1}{\kappa}\sum_{j\neq i}\frac{p_{i}\cdot p_{j}}{z_{i}-z_{j}}I_{0}.\label{eq:tachyon-correlator}
\end{equation}
(Assuming that we identify $\alpha'=1/\kappa$.) Starting with $\phi_{0}(z_{1}^{0},z_{2}^{0},\dots,z_{n}^{0})$
at a specific set of $z_{i}$'s the KZ connection uniquely determines
the value $\phi(z'_{1},z'_{2},\dots,z'_{n})$ through parallel transport,
and therefore $I_{0}(z'_{1},z'_{2},\dots,z'_{n})$ at any set of $z_{i}$'s.
Especially when the number of punctures $n$ is restricted to $2$
we see the bilinear form $\Bigl\langle v_{p},\,v_{s}\Bigr\rangle:=\Bigl\langle V_{-p}(z_{2})\,V_{s}(z_{1})\Bigr\rangle$
for all $(z_{1},z_{2})$ including the asymptotics $(0,\infty)$ are
related to each other in the same manner.

In the case where one gluon is present, suppose instead of direct
substitution with a gluon vertex operator $V_{p_{1}}^{\text{gluon}}(z_{1})=\epsilon\cdot\dot{X}e^{ip_{1}\cdot X(z_{1})}$
we choose to represent gluon by the Del Giudice-Di Vecchia-Fubini
(DDF)  constructed vertex \cite{DelGiudice:1971yjh},
\begin{equation}
V_{p_{1}}^{\text{gluon}}(z_{1})=\int_{\ell_{1}}dt\,e^{ip_{0}\cdot X(z_{1})}S_{k_{0}}(t),\label{eq:ddf}
\end{equation}
where $S_{k_{0}}(t)=\epsilon\cdot\dot{X}e^{ik_{0}\cdot X(t)}$. The
polarisation is taken to be in the orthogonal direction $\epsilon\cdot p_{0}=0$,
and $p_{0}+k_{0}=p_{1}$ is the original gluon momentum. Note the
settings of DDF demands $\alpha'p_{0}\cdot k_{0}=-1$ so that equation
(\ref{eq:ddf}) can be regarded as a special case of the screening
$E_{k_{o}}$ acting on a module $V_{p_{0}}(z_{1})$, where
the branch cut vanishes because of the integer exponent so that
 we can
safely close the C-shaped contour (since now its boundaries are
on the same sheet) to form a closed loop $\ell_{1}$
around $z_{1}$. 
From this perspective the one gluon correlator is the
pairing of the cycle $\ell_{1}$ with a $1$-form derived from OPE,
which in turn can be easily identified term by term with the $1$-forms
generated by KZ coefficients $I_{(0,\dots,1,\dots,0)}=\int_{\ell_{1}}dt\frac{1}{t-z_{i}}\Phi_{\kappa}$.
\begin{equation}
\Bigl\langle 0 \Bigr|V_{p_{n}}(z_{n})\dots V_{p_{2}}(z_{2})\left(E_{k_{0}}V_{p_{0}}(z_{1})\right)\Bigl| 0 \Bigr\rangle=\sum_{i=2}^{n}\epsilon\cdot p_{i}\,I_{\underset{i\text{-th entry}}{(0,\dots,1,\dots,0)}},\label{eq:gluon-correlator}
\end{equation}
In light of the $1$-level lowered KZ solution is given by the following
sum of tensor modules,
\begin{align}
\phi_{1}= & I_{(1,0,\dots,0)}\,f_{k_{0}}\tilde{v}_{p_{0}}\otimes\tilde{v}_{p_{2}}\otimes\dots\otimes\tilde{v}_{p_{n}}+I_{(0,1,\dots,0)}\,\tilde{v}_{p_{0}}\otimes f_{k_{0}}\tilde{v}_{p_{2}}\otimes\dots\otimes\tilde{v}_{p_{n}}+\dots\nonumber \\
 & +I_{(0,0,\dots,1)}\,\tilde{v}_{p_{0}}\otimes\tilde{v}_{p_{2}}\otimes\dots\otimes f_{k_{0}}\tilde{v}_{p_{n}},
\end{align}
the one gluon correlator can be expressed as $\phi_{1}$ projected
onto a dual vector $0\,w_{(1,0,\dots,0)}+\epsilon\cdot p_{2}\,w_{(0,1,\dots,0)}+\dots+\epsilon\cdot p_{n}\,w_{(0,\dots,0,1)}$.
The action of Casimir $\Omega_{ij}$ can be again read off directly
from the module, yielding a slightly more complicated set of differential
equations that relates correlators at different $z_{i}$'s. As a quick
consistency check of the relations just described, recall that in
the zero string tension limit $\alpha'\rightarrow\infty$, the correlator
should only have support on the Gross-Mende saddle points \cite{Gross:1987ar}. 
Suppose if we fix the values of $(p_{0},p_{2},\dots,p_{n})$ while maintaining
the condition $\alpha'p_{0}\cdot k_{0}=-1$, so that $k_{0}\sim 1/\alpha'\rightarrow0$
and the action of root vector $e_{k_{0}}$ on modules becomes negligible,
the KZ equations of $\phi_{1}$ implies
\begin{equation}
\frac{\partial}{\partial z_{i}}\left\langle V_{p_{n}}(z_{n})\dots V_{p_{2}}(z_{2})V_{p_{1}}^{\text{gluon}}(z_{1})\right\rangle =\frac{1}{\kappa}\sum_{j\neq i}\frac{p_{i}\cdot p_{j}}{z_{i}-z_{j}}\,\left\langle V_{p_{n}}(z_{n})\dots V_{p_{2}}(z_{2})V_{p_{1}}^{\text{gluon}}(z_{1})\right\rangle ,
\end{equation}
therefore we see that scattering equations indeed must be satisfied
if the $z_{i}$'s are to localise. It is straightforward to generalise
the above reasoning to incorporate more higher modes in the string
spectrum, for example an $n$-gluon correlator is given by the pairing
of an $n$-cycle with the $n$-form derived from the OPE, whereas
the $n$-cycle, when visualised on the punctured plane, is given by
the $n$ independent closed loops $\ell_{1}$, $\dots$, $\ell_{n}$
surrounding each puncture $z_{i}$. The corresponding $n$-form, on
the other hand, can be spanned by the $n$-forms generated by KZ coefficients
through straightforward term by term identifications.

\subsection{KZ solutions and Z-amplitudes}
\label{sec:z-amplitudes}

We return to amplitudes and numerators. In the previous section we
directly identified the KZ coefficients needed to span a correlator.
Generically a correspondence is known as the Drinfeld-Kohno theorem
\cite{Kohno87,Drinfeld:1989st} which identifies given quantum
algebra $U_{q}(g)$ behaviour with the monodromy of $\phi$'s, which
lives in the representation space of classical Kac-Moody algebra $g$
with the same roots as $U_{q}(g)$. In view of the discussions in
section \ref{sec:bcj-numerators} we see that an $SL(2,\mathbb{R})$
fixed $n$-point amplitude or numerator is described by the QUEA with
simple roots $\left\{ k_{2},k_{3},...,k_{n-2}\right\} $ read off
from its external legs as in (\ref{eq:screeningmodule}), it is therefore
natural to look for KZ coefficients in the $\sum_{i=2}^{n-2}k_{i}$-
weight lowered level subspace of the classical tensor module $\tilde{V}_{k_{1}}\otimes\tilde{V}_{k_{n-1}}$
.

For example at four points, suppose if we fix $(z_{1},z_{3},z_{4})$
at two arbitrary points and infinity respectively, the associated
algebra is then given by (\ref{eq:kac-moody-cr}) with only one (simple)
root $k_{2}$. For this algebra, the Casimir operator (\ref{eq:casimir}) reduces to:
\begin{equation}
\Omega=\sum_{i=0}^{D-1}h_i \otimes h_{i}+f_{k_2}\otimes e_{k_2}+e_{k_2}\otimes f_{k_2}\label{eq:4ptcasimir}
\end{equation}
The KZ equation is solved on the maps from $(z_{1},z_{3})\in\Bigl\{\mathbb{C}^{2}\Bigr|z_{1}\neq z_{3}\Bigr\}$
to vectors in $\tilde{V}_{k_{1}}\otimes\tilde{V}_{k_{3}}$, which
has the following form.
\begin{equation}
\phi_{4}=I_{(\left\{ k_{2}\right\} ,\emptyset)}\,f_{k_{2}}\tilde{v}_{k_{1}}\otimes\tilde{v}_{k_{3}}+I_{(\emptyset,\left\{ k_{2}\right\} )}\,\tilde{v}_{k_{1}}\otimes f_{k_{2}}\tilde{v}_{k_{3}}.
\end{equation}
Here we denote the coefficients as $I_{(\left\{ k_{2}\right\} ,\emptyset)}$
to emphasise generically they should be labeled by an ordered set
that clarifies in which order the root vector $f_{k_{i}}$'s are applied
to the corresponding tensor vector. In terms of this notation the
solution to the KZ equations is given by the following.
\begin{equation}
I_{(\left\{ k_{2}\right\} ,\emptyset)}=\int_{\gamma}dt\frac{1}{t-z_{1}}\Phi_{\kappa,},\,I_{(\emptyset,\left\{ k_{2}\right\} )}=\int_{\gamma}dt\frac{1}{z_{3}-t}\Phi_{\kappa},
\end{equation}
where
\begin{equation}
\Phi_{\kappa}=\left(t-z_{1}\right)^{k_{2}\cdot k_{1}/\kappa}\left(z_{3}-t\right)^{k_{3}\cdot k_{2}/\kappa}\left(z_{3}-z_{1}\right)^{k_{1}\cdot k_{3}/\kappa},
\end{equation}
and $\gamma$ is any closed contour, for example the
Pochhammer encircling $z_{1}$ and $z_{3}$. A $Z$-theory amplitude
$A_{P}(1,2,3,4)$ \cite{Carrasco:2016ldy,
Carrasco:2016ygv,Mafra:2016mcc} for example is known to be expressible
(up to proportionality factors produced when translating between ordered
integrals and Pochammer) as the linear combination $I_{(\left\{ k_{2}\right\} ,\emptyset)}+I_{(\emptyset,\left\{ k_{2}\right\} )}$
and therefore satisfies the KZ equations, in the sense that it can
be expressed as scalar product of $\phi_{4}$ and
a $z_{i}$ independent dual module, assuming the orthonormal duals
to the two basis tensor vectors in (\ref{eq:first}) are $w_{\left(\left\{ k_{2}\right\} ,\emptyset\right)}$
and $w_{\left(\emptyset,\left\{ k_{2}\right\} \right)}$.
\begin{equation}
A_{P}(1,2,3,4)=(w_{\left(\left\{ k_{2}\right\} ,\emptyset\right)}+w_{\left(\emptyset,\left\{ k_{2}\right\} \right)},\phi_{4})=\int_{\gamma_{P}}dt\,\frac{z_{3}-z_{1}}{t-z_{1}}\,\frac{1}{z_{3}-t}\,\Phi_{\kappa,\left\{ \alpha_{1}\right\} }.
\end{equation}
Similarly for the five-point Z-amplitude, the algebra contains two simple
roots $k_{2}$ and $k_{3}$, and the target space of KZ equation is
the $\sum_{i=2}^{3}k_{i}$-lowered level of $\tilde{V}_{k_{1}}\otimes\tilde{V}_{k_{4}}$.
(Namely, the weight $\sum_{i=1}^{4}k_{i}$ subspace.)  For generic value of $k_2$ and $k_3$ the Casimir operator in (\ref{eq:casimir}) is a infinite sum, as any positive integer sum of $k_2$ and $k_3$ will appear as a root in the summation. However, in the space of vectors of the form: 
\begin{align}
\phi_{5} & =I_{(\left\{ k_{2},k_{3}\right\} ,\emptyset)}f_{k_{2}}f_{k_{3}}v_{k_{1}}\otimes v_{k_{4}}+I_{(\left\{ k_{2}\right\} ,\left\{ k_{3}\right\} )}f_{k_{2}}v_{k_{1}}\otimes f_{k_{3}}v_{k_{3}}\nonumber \\
 & +I_{(\emptyset,\left\{ k_{2}k_{3}\right\} )}v_{k_{1}}\otimes f_{k_{2}}f_{k_{3}}v_{k_{4}}+(k_{2}\longleftrightarrow k_{3}).
\end{align}
the Casimir operator is effectively
\begin{align}
\Omega&=\sum_{i=0}^{D}h_{i}\otimes h_{i}+\sum_{i=2,3}\left(e_{k_{i}}\otimes f_{k_{i}}+e_{k_{i}}\otimes f_{k_{i}}\right)\\\nonumber
&+\frac{1}{k_{2}\cdot k_{3}}\left[e_{k_{2}},e_{k_{3}}\right]\otimes\left[f_{k_{2}},f_{k_{3}}\right]+\frac{1}{k_{2}\cdot k_{3}}\left[f_{k_{2}},f_{k_{3}}\right]\otimes\left[e_{k_{2}},e_{k_{3}}\right],
\end{align}
as all terms involving higher orders of $e_{k_i}$ will vanish in the subspace we are considering here. 
The solution for the first three $I$'s takes the
following form.
\begin{align}
I_{(\left\{ k_{2},k_{3}\right\} ,\emptyset)} & =\int_{\gamma}dt_{2}dt_{3}\frac{1}{t_{3}-t_{2}}\frac{1}{t_{3}-z_{1}}\Phi_{\kappa},\\
I_{(\left\{ k_{2}\right\} ,\left\{ k_{3}\right\} )} & =\int_{\gamma}dt_{3}dt_{3}\frac{1}{t_{2}-z_{1}}\frac{1}{z_{4}-t_{3}}\Phi_{\kappa},\\
I_{(\emptyset,\left\{ k_{2},k_{3}\right\} )} & =\int_{\gamma}dt_{2}dt_{3}\frac{1}{t_{3}-t_{2}}\frac{1}{z_{4}-t_{3}}\Phi_{\kappa},
\end{align}
with the Koba-Nielsen factor $\Phi_{\kappa}$ given
by
\begin{equation}
\Phi_{\kappa}=\left(z_{4}-z_{1}\right)^{k_{1}\cdot k_{4}/\kappa}\left(t_{3}-t_{2}\right)^{k_{2}\cdot k_{3}/\kappa}\prod_{i=2,3}\left(t_{i}-z_{1}\right)^{k_{i}\cdot k_{1}/\kappa}\left(z_{4}-t_{i}\right)^{k_{i}\cdot k_{4}/\kappa}.
\end{equation}
The rest three coefficients can be obtained by permutations of $t_{2}$
and $t_{3}$. Note that the solution $\phi_{5}$ explicitly depends
on the homology class of the integration domain. For
this reason we shall write the coefficients $I$ and $\phi_{5}$
as a function of the homology class $\left[\gamma\right]$ of $\gamma$. 

Similarly to the four-point case, the $Z$-amplitude
at 5 points can be written as (signed) sum of the
KZ coefficients $I$'s, for example,
\begin{equation}
A_{P}\left(1,2,3,4,5\right)=I{}_{\left\{ k_3,k_2\right\} ,\emptyset}\left(\left[\gamma_{P}\right]\right)+I_{\left\{k_2\right\} ,\left\{ k_3\right\} }\left(\left[\gamma_{P}\right]\right)+I_{\emptyset,\left\{ k_2,k_3\right\} }\left(\left[\gamma_{P}\right]\right)
\end{equation}
where the integration domain $\gamma_{P}$ is the $2$-dimension
generalization of the Pochammer contour, which is a $2$-cycle in
$\mathcal{C}_{2,2}(z)=\Bigl\{ t\in\mathbb{C}^{2}\Bigr|t_{2},t_{3}\neq z_{1},z_{4},\,t_{2}\neq t_{3}\Bigr\}$
that can be identify with the integration domains
correspond to the relative order $P\left[2,3\right]$: $z_{1}\leq t_{2}\leq t_{3}\leq z_{4}$
for $P\left[2,3\right]=\left\{ 2,3\right\} $, and $z_{1}\leq t_{3}\leq t_{2}\leq z_{4}$
for $P\left[2,3\right]=\left\{ 3,2\right\} $. Therefore the $Z$-theory
amplitudes $A_{P}\left(1,Q\left\{ 2,3\right\} ,4,5\right)$ can be
constructed from $\phi_{5}$ in the following way.

\begin{equation}
A_{P}\left(1,Q\left\{ 2,3\right\} ,4,5\right)=(w_{Q\left\{ 2,3\right\} },\phi_{5}\left(\left[\gamma_{P\left\{ 2,3\right\} }\right]\right))
\end{equation}
where the two dual vectors read 
\begin{equation}
w_{\left\{ 2,3\right\} }=\left(f_{k_{3}}f_{k_{2}}v_{k_{1}}\otimes v_{k_{4}}\right)^{*}+\left(f_{k_{2}}v_{k_{1}}\otimes f_{k_{3}}v_{k_{3}}\right)^{*}+\left(v_{k_{1}}\otimes f_{k_{2}}f_{k_{3}}v_{k_{4}}\right)^{*}
\end{equation}
and 
\begin{equation}
w_{\left\{ 3,2\right\} }=\left(f_{k_{2}}f_{k_{3}}v_{k_{1}}\otimes v_{k_{4}}\right)^{*}+\left(f_{k_{3}}v_{k_{1}}\otimes f_{k_{2}}v_{k_{3}}\right)^{*}+\left(v_{k_{1}}\otimes f_{k_{3}}f_{k_{2}}v_{k_{4}}\right)^{*}
\end{equation}
Notably, the KZ equation was introduced earlier to
the study of string amplitudes by Broedel, Schlotterer, Stieberger
and Terasoma in \cite{Broedel:2013aza} to construct
the Drinfeld associator that relates higher point $Z$-integrals
to lower point ones. (See also \cite{Vanhove:2018elu,Puhlfuerst:2015gta,
Kaderli:2019dny}.) The
KZ equation used in \cite{Broedel:2013aza} is a normalised 
version of KZ equation from Kac-Moody algebra with simple roots $\left\{ k_{2},k_{3},...,k_{n-2}\right\} $
solving for $\phi$ on the weigh-$\sum_{i=2}^{n-2}k_{i}$ lowered
submodule a tri-tensor space $\tilde{V}_{k_{1}}\otimes\tilde{V}_{k_{0}}\otimes\tilde{V}_{k_{n-1}}$
with coordinates $\left(z_{1},z_{0},z_{n-1}\right)$.
\begin{equation}
\frac{\partial}{\partial z_{0}}\phi=\left(\frac{\Omega_{1,0}}{z_{0}}+\frac{\Omega_{0,n-1}}{z_{0}-z_{n-1}}\right)\phi\label{eq:z0KZ}
\end{equation}
 matrix representation of (\ref{eq:z0KZ}) acting on $Z$-integrals
can be used to build the Drinfeld associator. And
the specific form of matrices used in \cite{Broedel:2013aza}
can be achieved by suitable linear transformations.
Note that when the basis of twisted cocycles is already given, the matrix representation of KZ equations as differential relation for twisted cocycles can be directly written down  by taking derivatives, removing exact terms and expanding on  basis. Such matrix representations have been discussed in detail by Mizera in \cite{Mizera:2019gea}. 


\section{Conclusions}
\label{sec:conclusions}

In this paper we showed, with the help of screening vertex operators,
that the string generalisation of the BCJ numerators previously derived
in \cite{BjerrumBohr:2010hn,Fu:2018hpu} have a natural quantum group explanation.
The associated algebra structure depends on the specific root system,
which in turn is entirely defined by external leg momenta. The definition
of a screening involves a string vertex operator followed by a contour
integration. For this setting to be interpreted as a representation
of the quantum group, a screening operator only needs to create a
$1$-chain on the punctured plane $\mathbb{C}-\{z_{1},z_{2},\dots,z_{n}\}$
while various choices for the vertex operator in the string spectrum
leads to different cohomology contents and corresponds to different
representations. Generically the representation depends on string
modes as well as on the exact location $z_{i}$ of the modules. We
showed that modules built from the same string modes (and therefore
lead to the same cohomology structures) but located at different $z_{i}$'s
are related to each other by the flat KZ connection. From this perspective
the action of a universal $R$-matrix has an explicit graphical interpretation
as the braiding of two punctures. 
In other words, quantum algebraic structure of string amplitude can be 
represented by sections of a local system over configuration spaces 
of $z_{i}$s, with modules of the kinematic algebra as its fibre and KZ 
connection as its flat connection. This local system can be isomorphically 
mapped to the local system used in discussing twisted homology for string 
amplitude, with each element in the module mapped to a class of twisted 
forms and the KZ connection mapped to the Gauss-Manin connection. 
In fact this identification is known to the quantum groups community
as part of a broader  structure built from hyperplane arrangements.
While recent years have definitely witnessed substantial progress 
already made from analytic and geometry observations, we feel optimistic
that more can be added along this direction.

Note particularly that BCJ numerator has been calculated in the
context of field theory using fusion rules in \cite{Chen:2019ywi}
up to the next-to-MHV level. At the momentum of writing it is not 
clear whether the field theory results are identical to those derived 
from the screenings. It would be interesting to see if any of the two 
could be used to improve calculation efficiency.

Finally we would like to briefly remark on a subtle, but important
issue regarding screening representation of the quantum algebra. The
vast majority of the discussions in the literature start with tachyon
vertex operators as the representation for simple root vectors. For
$sl(N)$ this construction is known to be particularly simple,
as the root system $k_{i}\cdot k_{j}=\pm1,\,\pm2$ combined with residue
theory ensures that all root vectors are tachyons. On the other hand
for root systems other than $sl(N)$ the screenings for non-simple
root are known to inevitably involve various string excitation modes
or even an infinite expansion such as (\ref{eq:alpha-beta-ope}).
A question naturally arises is how to interpret gluon in this quantum
algebraic setting. From a pragmatic viewpoint, the integrand of screening
only needs to satisfy the braiding relations (\ref{eq:braiding-1}),
(\ref{eq:braiding-2}) for the purpose of serving as a representation
of the quantum algebra, and indeed this was the perspective taken
for example in \cite{Lentner:2017dkg}. An alternative interpretation can
be obtained through RNS strings. Note that a $bc$ system-like screening
was discussed in \cite{Etingof:1998ru} and can be used to explain
gluons subject to some modest notation matching. In addition the gluon
can be understood as a composite object built from tachyons only,
in a manner similar to the DDF but slightly modified approach, or
as various residues similar to the viewpoint taken in \cite{Frenkel:1980rn}.
We leave this part of the discussion to future work as it involves
much details and is perhaps better expanded in a separate paper.



\section*{Acknowledgements}

We would like to thank Song He, Pei-Ming Ho, Yu-Tin Huang, 
Kirill Krasnov and Pierre Vanhove 
for valuable discussions and comments during various stages of this work. 
We would also like to thank the anonymous referee for his/her thorough review
and in particular the very much encouraging suggestions for improving this paper.
CF would like to thank National Taiwan University for
their hospitality, where a substantial part of this work was made.
CF is supported by the Fundamental Research Funds for the Central
Universities (GK201803018). YW is supported by MoST grant 106-2811-M-002-196.


\end{document}